\documentclass[11pt]{article}
\usepackage{amsmath,amssymb,color,graphics,epsfig}
\usepackage{graphicx}
\usepackage{subfigure}
\usepackage{amsfonts}
\newcommand{\be}{\begin{equation}}
\newcommand{\ee}{\end{equation}}
\newcommand{\bea}{\setlength\arraycolsep{2pt} \begin{eqnarray}}
\newcommand{\eea}{\end{eqnarray}}
\newcommand{\nn}{\nonumber}
\newcommand{\cOR}{{\mathcal O}_\text{R}}

\newcommand{\GH}{\text{GH}}
\newcommand{\bulk}{\text{bulk}}

\newcommand{\joint}{\text{joint}}

\newcommand{\UR}{U_\text{R}}

\newcommand{\tR}{t_\text{R}}
\newcommand{\tL}{t_\text{L}}
\newcommand{\vR}{v_\text{R}}
\newcommand{\vL}{v_\text{L}}

\newcommand{\tLc}[1]{t_{\text{L},c{#1}}}
\newcommand{\tRc}[1]{t_{\text{R},c{#1}}}
\newcommand{\ro}{{r}^{*}_1}
\newcommand{\rt}{{r}^{*}_2}

\newcommand{\oor}{\text{\uppercase\expandafter{\romannumeral1}}}
\newcommand{\tr}{\text{\uppercase\expandafter{\romannumeral2}}}
\newcommand{\ttr}{\text{\uppercase\expandafter{\romannumeral3}}}
\newcommand{\fr}{\text{\uppercase\expandafter{\romannumeral4}}}

\def\ft#1#2{{\textstyle{\frac{\scriptstyle #1}{\scriptstyle #2} } }}
\def\fft#1#2{{\frac{#1}{#2}}}

\def\0{{\sst{(0)}}}
\def\1{{\sst{(1)}}}
\def\2{{\sst{(2)}}}
\def\3{{\sst{(3)}}}
\def\4{{\sst{(4)}}}
\def\5{{\sst{(5)}}}
\def\6{{\sst{(6)}}}
\def\7{{\sst{(7)}}}
\def\8{{\sst{(8)}}}
\def\sst#1{{\scriptscriptstyle #1}}

\begin{document}

\begin{flushright}
%\hfill{KIAS-P12028}
 %\hfill{
%\bf hep-th/yymmnnn}
\end{flushright}

\vspace{25pt}
\begin{center}
{\large {\bf Holographic complexity under a global quantum quench}}

\vspace{10pt}
 Zhong-Ying Fan$^1$, Minyong Guo$^{2,3}$\\

\vspace{10pt}
$^1${ Center for Astrophysics, School of Physics and Electronic Engineering, \\
 Guangzhou University, Guangzhou 510006, China }\\
 $^2${ Department of Physics, Beijing Normal University, \\
 Beijing 100875,  P. R. China}\\
$^3${ Perimeter Institute for Theoretical Physics \\Waterloo, Ontario N2L 2Y5, Canada\\}
\smallskip
%{\it $^{2}$Department of Physics and State Key Laboratory of Nuclear Physics and Technology,\\}
%{\it Peking University, No.5 Yiheyuan Rd, Beijing 100871, P.R. China\\}
%\smallskip
%{\it $^{3}$Collaborative Innovation Center of Quantum Matter, No.5 Yiheyuan Rd,\\}
%{\it  Beijing 100871, P. R. China\\}

\vspace{40pt}

\underline{ABSTRACT}
\end{center}
There are several different proposals, relating holographic complexity to the gravitational objects defined on the Wheeler-DeWitt patch. In this paper, we investigate the evolution of complexity following a global quantum quench for these proposals.
We find that surprisingly they all reproduce known properties of complexity, such as the switchback effect. However, each of these proposals also has its own characteristic features during the dynamical evolution, which may serve as a powerful tool to distinguish the various holographic duals of complexity.

\vfill {\footnotesize  Email: fanzhy@gzhu.edu.cn,\quad guominyong@gmail.com\,.}

\thispagestyle{empty}

\pagebreak

\tableofcontents
\addtocontents{toc}{\protect\setcounter{tocdepth}{2}}

%%%%%%%%%%%%%%%%%%%%%%%%%%%%%%%%%%%%%%%%

%\newpage
%%%%%%%%%%%%%%%%%%%%%%%%%%%%%%%%%%%%%%%%

%\vspace{2cm}

\section{Introduction}
The recent developments in AdS/CFT correspondence reveal that quantum information theory becomes a new bridge to connect quantum gravity, quantum field theory and condensed matter physics. Thereinto, quantum computational complexity (or complexity in brief) plays a crucial role to measure the distance between two quantum states. It is defined by the minimum number of gates needed to prepare a particular target state from a certain reference state in quantum mechanics \cite{Aa, Wa}. However, when applying this idea to quantum field theory, it is challenged since there is not a concrete definition of complexity for systems with infinitely many degrees of freedoms.

Recently, inspired by the geometric approach developed by Nielsen et.al \cite{Ni,NDGD,AD}, a quantum circuit model for the preparation of Gaussian states for a free scalar field theory was first proposed in \cite{JM1}. The idea is soon generalized to the femionic theory \cite{HM1,RKK}. Roughly at the same time, there appears a similar model, which was established on Fubini-Study metric \cite{CPMP}. This definition is refered to {\it circuit complexity} in literature. Although it is introduced for a free field theory, it was shown \cite{JM1} that it shares a number of qualitative features with holographic complexity. From then on, the circuit complexity has been widely studied. For example, the circuit complexity for coherent state, thermofield double states and coherent thermofield double states was studied in \cite{GHMR,CEHH} and \cite{FG2}, respectively. Along this line, see more examples in \cite{Bhattacharyya:2018bbv,JSY,WFA,Camargo:2018eof,Ali:2018fcz}. There are also different definitions of complexity for quantum field theories, such as the path-integral complexity \cite{CKMT,Bhattacharyya:2018wym} and the complexity of operator using the Finsler geometry \cite{YANZ,Yang:2018tpo}. In any event, the aim of these studies is to well understand the complexity in quantum field theory and find a comparison with the results in holography.

However, the notion of complexity is studied even more earlier than its counterpart in field theories. Based on the fact that the volume of Einstein-Rosen-Bridge (ERB) grows linearly with time for a very long time, the authors in \cite{MS} argued that the entanglement is not enough and the complexity is the right quantity to capture the information in the dual CFT since the former is saturated in thermal equilibrium \cite{Su1,Su2}. The first gravity dual of complexity was first postulated in \cite{SS}, dubbed by ``Complexity=Volume" (CV) duality. It states that the complexity is dual to the volume of the maximal spacelike hypersurface crossing the Einstein-Rosen bridge
\be \mathcal{C}\sim \fft{V_{max}}{G\ell_0} \,,\ee
where $\ell_0$ is some length scale. The main advantage of this conjecture is that it correctly captures the linear growth of complexity at late time. However, the disadvantage is also obvious since the length scale $\ell_0$ has to be chosen by hand. In order to avoid this shortcoming, a new conjecture was proposed in \cite{BRSSZ1,BRSSZ2}, dubbed by ``Complexity=Action" (CA) duality. It states that the complexity is dual to the gravitational action evaluated on a certain region of spacetime, called Wheeler-DeWitt (WDW) patch, which is defined as the causal development of a bulk Cauchy surface that is anchored on the boundary state. One has
\begin{equation}
\mathcal{C}=\frac{S_{grav}}{\pi\hbar}\,,
\end{equation}
where the overall coefficient is fixed by requiring the rate of change of complexity saturates the conjectured Lloyd bound \cite{Lloyd} for a Schwarzschild black hole. However, the CA duality has drawbacks as well. As pointed out in \cite{BRSSZ1,BRSSZ2}, for a Reissner-Nordstr\o m (RN) black hole, the CA duality generally does not respect the Lloyd bound. This motivates us to search better holographic duals for complexity.

In \cite{CFN}, it was established that the spacetime volume of the WDW patch might be such a better dual of complexity, dubbed by ``Complexity=Volume 2.0" (CV-2) duality. One has
\begin{equation}
\mathcal{C}=\frac{1}{\pi\hbar}P(\mathrm{Spacetime ~Volume})\,,
\end{equation}
where $P=-\Lambda/8\pi G$ stands for the thermodynamic pressure. It was proved in \cite{CFN} that this new conjecture respects the Lloyd bound for a RN black hole (at least for the four dimensional solution).

Later on, inspired by this work and considering the physical relevance of thermodynamic volumes to holographic complexity, we propose \cite{FG} that the non-derivative action associated to the cosmological constant evaluated on the WDW patch is dual to the complexity of the boundary theory, dubbed by ``Complexity=Action 2.0" (CA-2) duality
\begin{equation}
\mathcal{C}=\frac{S_\Lambda}{\pi\hbar}\,.
\end{equation}
For example, for black holes in Einstein-Scalar gravities (\ref{ESG})
\be S_\Lambda=-\fft{1}{16\pi G}\int_{WDW}\mathrm{d}^n x \sqrt{-g}\,V_\Lambda(\phi) \,,\ee
 where $V_\Lambda(\phi)$ is a part of the scalar potential $V(\phi)$ that is linearly proportional to the cosmological constant. We prove that under reasonable assumptions, the rate of change of complexity is essentially equal to the thermodynamic pressure times the thermodynamic volume for a wide class of eternal black holes: $\dot{\mathcal{C}}=P V_{ther}/\pi\hbar$ or $\dot{\mathcal{C}}=P (V_+-V_-)/\pi\hbar$ for black holes with an inner horizon. We further show that in many cases, the CA duality does not respect the Lloyd bound whereas our CA-2 duality always does.

However, none of above conjectures respects the Lloyd bound for all the known black holes. Despite that the above new proposals show interesting properties of complexity as the CA duality, they do not really obey an universal upper bound but they do have an advantage of computational simplicity. In fact, the existence of an upper bound on computational speed, although easily accepted, is still a conjecture that is surprisingly hard to prove in the field of quantum computations. Recently, it was established in \cite{Jordan:2017vqh} that in certain cases, the computational clock speed could be arbitrarily large compared to the energy of the dual state. Hence, in order to derive an upper bound on computation, one may need consider additional conditions such as the information density and the information transmission speed. In this sense, the Lloyd bound may be insufficient to distinguish the various holographic duals of complexity.

In this paper, we investigate the evolution of holographic complexity following a global quantum quench for the three proposals CA/CA-2/CV-2 duality, for a number of hairy black holes, either onesided or two-sided, in Einstein-Scalar gravity. Some previous studies on this topic for the CA and CV duality can be found in a series of papers \cite{SS,BRSSZ2,SZ,MMo,RSS1,CMM1,CMM2,
Tanhayi:2018gcj,Ageev:2018nye,Jiangjie20181108}. We find that surprisingly, all these different proposals reproduce known properties of complexity, such as the switchback effect. Moreover, each of these proposals also has its own characteristic features in the quench process. For example, under a thermal quench, the complexity at the early time logarithmically increases/decreases for the CA/CV-2 duality while for the CA-2 duality, it grows linearly with time. This may serve as a powerful tool to distinguish the various holographic duals of complexity.

The paper is organized as follows. In section 2, we illustrate the gravity model, describing a global quantum quench and discuss the action principle for dynamical hairy black holes. In section 3, we study the evolution of complexity following a thermal quench, described by the formation of an onesided black hole. In section 4, we study the evolution of complexity for thermofield double states perturbed by a global shock wave. We conclude in section 5 and discuss some future directions.

\section{The model setup and action principle}

 In order to study the evolution of holographic complexity following a global quantum quench as general as possible, we consider generally static AdS black holes with spherical/hyperbolic/toric isometries. Here and in the subsequent sections, we follow closely the prescriptions established in \cite{CMM1,CMM2} and extend the discussions there to general shockwave geometries. The metric ansatz is given by
 \be ds^2=-h(r)dt^2+dr^2/f(r)+r^2 d\Omega_{n-2\,,k}^2 \,,\ee
 where $d\Omega_{n-2\,,k}^2$ is the metric of the $(n-2)$-dimensional subspace and $k=1\,,-1\,,0$ corresponds to spherical/hyperoblic/toric topologies. The AdS vacuum has $h=f=r^2\ell^{-2}+k$. For later purpose, it is more convenient for us to work in ingoing Eddington-Finkelstein like coordinate
 \be
ds^2=-h(r)dv^2+2w(r)dvdr+r^2d\Omega_{n-2\,,k}^2\,,
\ee
where $h(r)=w(r)^2f(r)$ and $v=t+r^*(r)$, $r^*$ is the tortoise coordinate defined as
  \be r^*(r)=-\int_{r}^{\infty}\fft{\mathrm{d}r}{w(r)f(r)} \,.\ee
Here we have chosen the integration constant such that $r^*(\infty)=0$. Then (global or local) quantum quenches in the boundary are holographically described by dynamical spacetimes which characterize the formation of black holes. For our purpose, we would like to study the evolution of complexity following a global quantum quench which acts on an AdS vacuum or an initial AdS black hole. The generally dynamical spacetime describing the quench could be written as
\be\label{dynamicansatz}
ds^2=-h(r\,,v)dv^2+2w(r\,,v)dvdr+r^2d\Omega_{n-2\,,k}^2\,.
\ee
However, to simplify the discussions, we model the quench process by the collapse of a thin-shell. In this case, the full dynamical spacetime is separated into two static patches. We have
\be
ds^2=\left\{
\begin{array}{ccc}
-h_1(r)dv^2+2w_1(r)dvdr+r^2d\Omega_{n-2\,,k}^2\,,\quad \mathrm{for}\quad v<v_s\,,\nn\\
-h_2(r)dv^2+2w_2(r)dvdr+r^2d\Omega_{n-2\,,k}^2\,,\quad \mathrm{for}\quad v>v_s\,,
\end{array}
\right.
\ee
where $v=v_s$ stands for the location of the thin-shell and $(h_1\,,w_1)$ and $(h_2\,,w_2)$ are the metric functions of the initial and final static black holes respectively. In this setup, the total energy of the state was shifted from $M_1$ (the initial black hole mass) to $M_2$ (the final black hole mass) at the critical time $v=v_s$. In other words, it can be characterized as
 \be M(v)=M_1\mathcal{H}(v_s-v)+M_2\mathcal{H}(v-v_s) \,,\ee
 where $\mathcal{H}$ stands for a Heaviside function. All the time dependence of the dynamical evolution and hence the evolution of complexity is attributed to the motion of the shell. Of course, there is something put by hand in this model, for example the canonical time $t$ will be discontinuous across the shell. This is obviously artificial since the original spacetime should be smooth everywhere. Nevertheless, the model essentially captures the properties of the time dependence of complexity following a global quantum quench, as will be shown later. Although technically more involved, it is straightforward to generalize the discussions to a collapse shell with a finite width or to general dynamical spacetimes.\\

\textbf{Action principle}\\

To compute complexity using CA duality, we shall first clarify what the thin-shell is composed of and its action principle. This is particularly important when the finite width effect is considered. It was shown in \cite{CMM1} that the shell could be null for a Vaidya black hole which has
\be w(r\,,v)=1 \,,\quad h(r\,,v)= r^2\ell^{-2}+k-\fft{16\pi}{(n-2)\omega_{n-2}}\fft{GM(v)}{r^{n-3}} \,.\ee
In fact, by plugging the metric functions into the Einstein tensor, one finds the only non-vanishing component is
\be G_{vv}=\fft{16\pi G\dot{M}(v) }{\omega_{n-2}\,r^{n-2}}\,\,.\ee
Hence to generate a Vaidya black hole, it is suffice to introduce a null fluid in the bulk which has an on-shell stress tensor $T=T_{vv}dvdv$. We refer the readers to \cite{CMM1} for detailed discussions on action principle of a null fluid. However, the construction there cannot be directly generalized to general dynamical black holes since the matter sources needed may not be null any longer. Interestingly, in recent years there are a list of exact dynamical hairy black holes having been successfully constructed in Einstein-Scalar gravity \cite{Zhang:2014sta,Lu:2014eta,Xu:2014xqa,Zhang:2014dfa,Fan:2015tua,Fan:2015ykb,Fan:2016yqv,Chen:2016qks}. The Lagrangian density is given by
\be\label{ESG} \mathcal{L}=R-\fft12\big(\partial\phi \big)^2-V(\phi) \,,\ee
where $V(\phi)$ is the scalar potential which has a small $\phi$ expansion as $V=-(n-1)(n-2)\ell^{-2}+\ft 12 m^2\phi^2+\cdots$. For certain type potentials, for example
\be\label{potential1} V=-2g^2\big(\cosh{\phi}+2 \big)-2\alpha^2 \Big( \phi\big(\cosh{\phi}+2 \big)-3\sinh{\phi}  \Big)\,, \ee
one can find an exact dynamical hairy black hole solution \cite{Zhang:2014sta}
\bea\label{zxf}
&&ds^2=-h dv^2+2w dr dv+r^2 d\Omega_{2\,,k}^2\,,\quad \phi=\log{\Big(1+\fft{a}{\rho} \Big)}\,,\quad w=\fft{2r}{\sqrt{4r^2+a^2}}\,,\nn\\
&&h=g^2\rho^2+(g^2-\alpha^2)a \rho+k-\ft 12 \alpha^2 a^2-\fft{a\dot a}{\sqrt{4r^2+a^2}}+\alpha^2\, r^2\log{\Big(1+\fft{a}{\rho} \Big)}\,,
\eea
where $\rho=\ft 12\big(\sqrt{4r^2+a^2}-a \big)$, $a=a(v)$ is the dynamical ``scalar charge". Here a dot denotes the derivative with respect to $v$ (it should not be confused with the derivative with respect to the boundary time for the rate of change of complexity). Substituting the above solution into the equations of motion, one finds the only non-vanishing term is
\be\label{zxfeq} E_{vv}=\fft{a\big(\ddot a+\alpha^2 a \dot a \big)}{2r^2} \,,\ee
which can be immediately solved as $a=q\tanh{\Big(\ft 12\alpha^2 q (v-v_i) \Big)}$ except for the static solution $a=q$.  The nontrivial dynamical solution describes the formation of a hairy black hole from an initial state at $v=v_i$. However, a feature that we are not appreciate is at the initial time the spacetime is pure AdS everywhere but has a naked singularity\footnote{The singularity is path-dependent in $(v\,,r)$ plane.} at the center. The problem would be cured if an additional null fluid was introduced to drive the bulk dynamical evolution. To benefit the readers, we present the mass and temperature of the solution in the static limit\footnote{Throughout this paper, we work in the unit of $c=1$, where $c$ is the speed of light.}
\be M=\fft{\omega_2 \alpha^2 q^3}{48\pi G}\,,\qquad T=\fft{\alpha^2 q^3-2k \sqrt{4r_h^2+q^2}}{8\pi r_h^2} \,,\ee
where $r_h$ is the radii of the event horizon.

Another example that we are interested in is \cite{Fan:2015ykb}
\bea\label{potential2} V&=&-\ft 12 \big(\cosh{\Phi}\big)^{\ft{ k_0^2}{2}}\Big(g^2-\alpha^2 \big(\sinh{\Phi}\big)^{\ft{2(n-1)}{n-2}} {}_2F_1\big(\ft{k_0^2}{8},\ft{n-1 }{n-2},\ft{2n-3}{n-2},-\sinh^2{\Phi}  \big)   \Big)\\
&\times &\Big(2(n-2)(n-1)-\ft 14 k_0^2(n-2)^2\tanh^2{\Phi} \Big)-\alpha^2 (n-2)(n-1)\big(\cosh{\Phi}\big)^{\ft{k_0^2}{4}}\big(\sinh{\Phi}\big)^{\ft{2(n-1)}{n-2}}\,,  \nn\label{dypotential}\eea
where $\Phi=\phi/k_0$. The solution reads
\bea\label{fan} ds^2&=&-hdv^2+2w dr dv+r^2 dx^idx^i\,,\qquad \phi=k_0\, \mathrm{arcsinh}\Big(\fft{a}{r} \Big)^{\fft{n-2}{2}}\,,\\
      w&=&\Big(1+\ft{a^{n-2}}{r^{n-2}} \Big)^{-\ft{k_0^2}{8}}\,,\qquad h=g^2 r^2-2w \dot{a}\Big(\fft{a}{r}\Big)^{n-3}-\alpha^2 a^2\Big(\fft{a}{r}\Big)^{n-3}
      {}_2F_1\Big(\ft{k_0^2}{8},\ft{n-1}{n-2},\ft{2n-3}{n-2},-\ft{a^{n-2}}{r^{n-2}} \Big)\,. \nn\label{dysol}\eea
Though the scalar potential and the metric function $h$ look a little complicated, their expressions will be greatly simplified for certain parameters. Again the only non-trivial equation of motion is
\be\label{faneq} E_{vv}=\fft{(n-2)a^{n-4}}{8r^{n-2}}\Big(8a\ddot a+\big(8(n-3)-(n-2)k_0^2 \big)\dot{a}^2+4(n-1)\alpha^2 a^2\dot a     \Big) \,.\ee
This equation can also be solved analytically, giving rise to several distinct classes of dynamical hairy black holes \cite{Fan:2015ykb}. In addition, in the static limit $a(+\infty)=q$, the mass and temperature of the solution are given by
\be M=\fft{(n-2)\omega_{n-2}\alpha^2 q^{n-1}}{16\pi G}\,,\qquad T=\fft{(n-1)\alpha^2 q^{n-1}}{4\pi r_h^{n-2}} \,.\ee

In fact, these exact dynamical solutions can be viewed as the collapse of a time-like thin-shell with an effective finite width. This is shown in Appendix A, where we discuss a special solution refered to {\it hairy BTZ black hole}. However, for holographic studies one may need the dynamical solutions with an arbitrarily desired scalar function $a=a(v)$. This is easily achieved by introducing an extra null fluid in the bulk. For the focus of this paper, we work in the thin-shell limit so that these two cases work equivalently well (see Appendix B for details) but important differences are expected when finite width effect is taken into account. Considering the fact that for certain cases the exact dynamical solutions, for example, the solution (\ref{zxf}), cannot be analytically continued to an AdS vacuum, we will always introduce a null fluid to the original theories.

To end this section, we would like to point out that for the three proposals of holographic complexity which relate the complexity to certain gravitational objects on the WDW patch, the contribution to the rate of change of complexity from a black hole singularity is rather subtle. Strictly speaking, it cannot be properly treated in physics without a well developed quantum field theory of gravity. However, in the semiclassical limit, it was found that the singularity indeed gives a finite contribution to the complexity growth rate for certain black holes (for example see \cite{BRSSZ1,BRSSZ2,CFN,FG}). For the hairy black holes we studied in this paper (the static limit of (\ref{zxf}) and (\ref{fan})), we find that this is also true for appropriately chosen parameters, as will be shown in the subsequent sections.

\section{Complexity following a thermal quench}
In this section, we study the evolution of complexity following a special case of global quenches: the thermal quench, which is described by the formation of an one-sided AdS black hole (see the Penrose-like diagram in Fig.\ref{fig3}). To calculate the rate of change of complexity for the different proposals, we need calculate the dual ``actions" defined on the WDW patch. We focus on discussing the CA duality since the remaining two cases involve only bulk terms which are included in the former as well.
\subsection{Gravitational action}
The total gravitational action that is relevant to us is
\bea
S_{grav}&=&\fft{1}{16\pi G}\int_{\mathcal{M}}\mathrm{d}^{n}x\,\sqrt{-g}\,\mathcal{L}+\fft{1}{8\pi G}\int_{r=\epsilon}\mathrm{d}^{n-1}x\,\sqrt{|\gamma|}\,K\nn\\
&&+\fft{1}{8\pi G}\int_{\Sigma}\mathrm{d}^{n-2}x\,\sqrt{\sigma}\,a+\fft{1}{8\pi G}\int_{\mathcal{B}}d\lambda\mathrm{d}^{n-2}\theta\,\sqrt{|\gamma_N|}\,\kappa\nn\\
&&+\fft{1}{8\pi G}\int_{\mathcal{B}} d\lambda d^{n-2}\theta \sqrt{|\gamma_{\mathcal{B}}|}\,\Theta \log{\big(\ell_{ct}\Theta \big)}+S_{shell}\,,
\eea
where the first is the bulk action of the WDW patch and the second is Gibbons-Hawking (GH) surface term at the future singularity. These two terms will be treated carefully in subsection \ref{onesidecomplexity}. The third term is corner contributions, originating from the joints at which two null boundaries intersect. In particular, the joints on the shell nontrivially contribute to the time dependence of complexity. The result is given in Eq.(\ref{onesidejointaction}).

\begin{figure}
  \centering
  	\includegraphics[width=200pt]{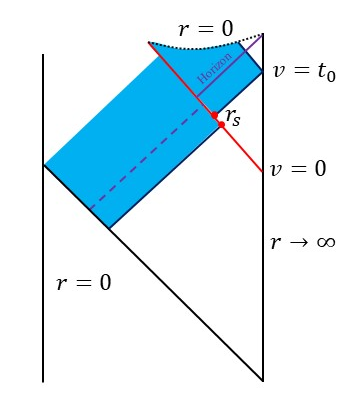}
  	\caption{The Penrose-like diagram for the formation of a AdS planar black hole.}
\label{fig3}\end{figure}

The fourth term is defined on the null boundaries associated to parameterization of their outward-directed null norms $k^\mu$. One has $k^\nu\nabla_\nu k^\mu=\kappa\,k^\mu$. At asymptotic AdS boundary, we impose normalization $k\cdot \partial_t=\pm \alpha$, where $\alpha$ is a positive constant and the sign ``$+(-)$" corresponds to the future (past) null boundaries. In particular, on the past null boundary in Fig.\ref{fig3}, we have
\bea
&&v>v_s\,,\quad k\cdot \partial_t=-\alpha\,,\nn\\
&&v<v_s\,,\quad k\cdot \partial_t=-\tilde\alpha\,,
\eea
where both $\alpha$ and $\tilde\alpha$ are positive constant but in general $\tilde\alpha\neq \alpha$ since the inner region does not extend to the asymptotic AdS boundary. Using this degrees of freedom, one can remove this part of action by choosing proper asymptotic normalization for the normal vectors (see Appendix B for details). One has
\be\label{normalization} \tilde\alpha=\fft{w_1(r_s)f_1(r_s)}{w_2(r_s)f_2(r_s)}\alpha \,,\ee
which leads to $\kappa=0$. In fact, the total gravitational action is independent of the reparameterization of the null boundaries when the counterterm action (the fifth term) is included \cite{Lehner:2016vdi}.

 The fifth term is a counterterm action that is introduced on null boundaries. This term was first introduced in \cite{Lehner:2016vdi}. Since it depends only on the intrinsic geometry of the null boundaries, it is unnecessary to the variational principle. However, it removes various ambiguities of the gravitational action associated to the null boundaries but it does not affect certain key results for the CA duality, such as the late time rate of change of complexity and the complexity of formation \cite{Chapman:2016hwi}. Interestingly, it was established in \cite{CMM1,CMM2} that this term plays an indispensable role to reproduce known properties of complexity following a global quantum quench. For example, the inclusion of this term is essential to produce the correct late time rate of change of complexity in the quench process. We will come to this point soon later.

 Finally, the last term $S_{shell}$ denotes the action on the thin-shell. For a Schwarzschild (or a Reissner-Nordstr\o m) black hole this term vanishes \cite{CMM1}. However, for general static black holes, it does not vanish any longer. We find
\be S_{shell}= \fft{\omega_{n-2}}{8\pi G} r_s^{n-2}\log{\Big( \fft{w_2(r_s)}{w_1(r_s)} \Big)}\,.\ee
This is proved in Appendix C where the readers can find more details. Furthermore, though this term is nonvanishing, it is actually cancelled by the corner contributions at $r=r_s$ on the past null boundary. Following the prescriptions in \cite{Lehner:2016vdi}, we have
\bea\label{onesidejointaction}
S_{joint}&=&\fft{1}{8\pi G}\int \mathrm{d}S (a_2-a_1)\nn\\
&=&\fft{\omega_{n-2}}{8\pi G}r_s^{n-2}\Big[\log{\big(\ft{2\alpha\beta}{h_{2}(r_s)} \big)}-\log{\big(\ft{2\tilde\alpha\beta}{h_{1}(r_s)} \big)} \Big]\nn\\
&=&-\fft{\omega_{n-2}}{8\pi G}r_s^{n-2}\log{\Big( \fft{w_{2}(r_s)}{w_1(r_s)} \Big)}\,,
\eea
where in the last line we have adopted the normalization Eq.(\ref{normalization}). Therefore,
\be S_{shell}+S_{joint}=0 \,.\ee
In short, for the CA duality, the total gravitational action that is relevant to the evolution of complexity reduces to the bulk term of the WDW patch, the GH surface term at the future singularity and the counterterm on the past null boundary
\be S_{grav}=S_{bulk}+S_{GH}+S_{ct}+\cdots \,,\ee
where the dotted terms do not have any time dependence. For the CA-2 and CV-2 duality, only the first bulk term survives with $\mathcal{L}=-V_{\Lambda}$ ( this is a part of the scalar potential relevant to the cosmological constant) for the CA-2 duality and $\mathcal{L}=-2\Lambda$ for the CV-2 duality.

Before moving on, we shall comment on the action growth at the late time limit for the CA proposal. In \cite{FG}, it was proved that for general static black holes, the late time growth of the action is essentially given by the bulk action evaluated in the black hole interior and the GH surface term evaluated both on the event horizon and at the future singularity. However, from the above result, we lose a piece of the action at the late time in the quench process: the GH surface term on the event horizon. Therefore, without including the counterterm action, the late time rate of change of complexity will be incorrect. To resolve this problem, we recall that the GH surface term on the event horizon gives
\bea
\fft{dS_{GH}}{dt_0}\Big|_{r=r_h}&=&\lim_{r\rightarrow r_h}\fft{\omega_{n-2}}{16\pi G}\,r^{n-2}w_{BH}f_{BH}\Big(\ft{h'_{BH}}{h_{BH}}+\ft{2(n-2)}{r}\Big)\nn\\
&=&\fft{\omega_{n-2}}{16\pi G}\,r_h^{n-2}w_{BH}(r_h)f_{BH}'(r_h)\nn\\
&=&T S \,,\eea
where $t_0$ stands for the boundary time and $T\,,S$ are the black hole temperature and entropy respectively. Fortunately, the counterterm action will contribute a same term at the late time (see Eq.(\ref{counterlate})) and thus cures the problem.

\subsection{Counterterm for null boundaries}\label{counteraction}
Considering the physical relevance of the counterterm action on the past null boundary, we shall first deal with it separately before moving to the complexity calculations. We rewrite it in the following
\be\label{sct} S_{ct}=\fft{1}{8\pi G}\int_{\mathcal{B}} d\lambda d^{n-2}\theta \sqrt{|\gamma_{\mathcal{B}}|}\,\Theta \log{\big(\ell_{ct}\Theta \big)} \,,\ee
where $\ell_{ct}$ is an arbitrary length scale and $\Theta $ is the scalar expansion defined by
\be \Theta=\partial_\lambda \log{\sqrt{|\gamma_{\mathcal{B}}|}}\,.\ee
Here $\lambda$ parameterizes the normal vectors. On the past null boundary, one has
\be\label{nullnorm} k=\partial_\lambda=H(r\,,v)\Big(\fft{2}{h(r\,,v)}\partial_v+\fft{1}{w(r\,,v)}\partial_r \Big) \,,\ee
 where $H$ is associated to the asymptotic normalization of the null norm. We have
\be \Theta=(n-2)\partial_\lambda\log{r}=\fft{(n-2)H(r\,,v) }{w(r\,,v) r} \,,\ee
leading to
\be S_{ct}=\fft{\omega_{n-2}}{8\pi G}\int_{r_{min}}^{r_{max}} \mathrm{d}\big(r^{n-2} \big)\log{\Big( \fft{(n-2)\ell_{ct} H }{w r} \Big)} \,,\ee
where $r_{max}$ stands for the UV cutoff and $r_{min}$ depends on the geometry we study. We place the thin shell at $r=r_s$ and take
\be H(r\,,v)=\alpha\mathcal{H}(r-r_s)+\tilde\alpha\Big( 1- \mathcal{H}(r-rs)\Big) \,,\ee
where $\mathcal{H}$ is the Heaviside function. Thus,
\bea\label{counter1}
S_{ct}&=&\fft{\omega_{n-2}}{8\pi G}\int_{r_{min}}^{r_s} \mathrm{d}\big(r^{n-2} \big)\log{\Big( \fft{(n-2)\ell_{ct} \tilde\alpha }{r} \Big)}\nn\\
&&+\fft{\omega_{n-2}}{8\pi G}\int_{r_s}^{r_{max}} \mathrm{d}\big(r^{n-2} \big)\log{\Big( \fft{(n-2)\ell_{ct} \alpha }{w_{BH}\, r} \Big)}\,.
\eea
To extract the time dependence, we isolate the $\log{w_{BH}}$ term
\be
S_{ct}=\fft{\omega_{n-2}}{8\pi G}\Big[r_s^{n-2}\log{\Big( \fft{\widetilde\alpha}{\alpha}\Big)}+\int_{r_{max}}^{r_s}\mathrm{d}\big(r^{n-2} \big)\log{w_{BH}}  \Big]+\cdots\,,
\ee
where the dotted terms are irrelevant to $r_s$ and hence do not have any nontrivial time dependence. Note that the first term in the square bracket is formally the same as the Vaidya black hole \cite{CMM1}. Using the normalization Eq.(\ref{normalization}), we deduce
\bea\label{counterterm}
S_{ct}
&=&\fft{\omega_{n-2}}{8\pi G}\Big[r_s^{n-2}\log{\Big(\fft{h_{vac}(r_s) }{f_{BH}(r_s)w_{BH}(r_s)}\Big)}+\int_{r_{max}}^{r_s}\mathrm{d}\big(r^{n-2} \big)\log{w_{BH}}\Big]+\cdots\,.
\eea
During the formation process, the location of the shell is a function of the boundary time $r_s=r_s(t_0)$, which is determined by Eq.(\ref{tbh2}).  Thus, we deduce
\bea\label{onesidecounter}
\fft{dS_{ct}}{dt_0}&=& \fft{\omega_{n-2}}{8\pi G}r_s^{n-2}\Big[ \fft{h_{vac}'(r_s)}{h_{vac}(r_s)}-\fft{f_{BH}'(r_s)}{f_{BH}(r_s)}-\fft{w_{BH}'(r_s)}{w_{BH}(r_s)}
+\fft{n-2}{r_s}\log{\Big(\fft{h_{vac}(r_s)}{f_{BH}(r_s)} \Big)}  \Big]\fft{dr_s}{dt_0}  \nn\\
&=&\fft{\omega_{n-2}}{16\pi G}r_s^{n-2}w_{BH}(r_s)\Big[f'_{BH}(r_s)-f_{BH}(r_s)\log'\Big(\fft{h_{vac}(r_s)}{w_{BH}(r_s)} \Big)\nn\\
&&\qquad\qquad\qquad\qquad\qquad\qquad\quad-\fft{(n-2)f_{BH}(r_s)}{r_s}\log{\Big(\fft{h_{vac}(r_s)}{f_{BH}(r_s)} \Big)} \Big] \,,
\eea
where in the second line, we have adopted the relation Eq.(\ref{tbh2}). It is immediately seen that in the late time limit $r_s\rightarrow r_h$, we have
\be\label{counterlate} \fft{dS_{ct}}{dt_0}=\fft{\omega_{n-2}}{16\pi G}r_h^{n-2}w_{BH}(r_h)f'_{BH}(r_h)=T S \,.\ee
It correctly captures the lost GH surface term on the event horizon. Furthermore, it was shown in \cite{CMM1,CMM2} that this term plays an essential role to reproduce known properties of complexity for the CA duality, for example the switchback effect for complexity of formation.

\subsection{The evolution of complexity}\label{onesidecomplexity}

Having established the gravitational actions on the WDW patch, we now would like to discuss the motion of the shell and calculate the complexity growth during the formation. The shell separates the spacetime into two static patches: the AdS black hole at $v>v_s$ and the AdS vacuum at $v<v_s$. For later convenience, we set $v_s=0$ and introduce the canonical time $t=v-r_*(r)$ and the retarded time $u=v-2r_*(r)$. Note that $t\,,u$ are discontinuous across the shell because of $r_*^{BH}(r_s)\neq r_*^{vac}(r_s)$. Initially the shell is anchored on the AdS boundary at some constant time $v=t_0$. When it collapses into the bulk, we have $u=t_0=v-2r_*^{BH}(r)$ along the past null boundary in the black hole region ($v>0\,,r>r_s$). This part of the WDW patch terminates at the future singularity at $v=t_0+2r_*^{BH}(0)$. At the location of the shell, one has
\be\label{tbh1} t_0+2r_{*}^{BH}(r_s)=0 \,.\ee
Taking a derivative with respect to the boundary time, one finds
\be\label{tbh2} \fft{dr_s}{dt_0}=-\ft 12 f_{BH}(r_s)w_{BH}(r_s)  \,.\ee
This describes the motion of the shell. On the other hand, in the AdS vacuum ($v<0\,,r<r_s$), we have $u=-2r_*^{vac}(r_s)=v-2r_*^{vac}(r)$. With these relations in hand, we are ready to compute the bulk gravitational action on the WDW patch as well as the GH surface term at the future singularity.

We deduce
\bea\label{bulkactionv2}
S_{bulk}&=&\fft{\omega_{n-2}}{16\pi G}\int_{0}^{r_s}\mathrm{d}r\Big(\sqrt{-\bar g}\mathcal{L}\Big)_{vac}\int_{2r_*^{vac}(r)-2r_*^{vac}(r_s)}^{0}dv \nn\\
  &&+\fft{\omega_{n-2}}{16\pi G}\int_{r_s}^{\infty}\mathrm{d}r\Big( \sqrt{-\bar g}\mathcal{L}\Big)_{BH}\int_{t_0+2r_*^{BH}(r)}^{t_0}dv\nn\\
   &&+\fft{\omega_{n-2}}{16\pi G}\int_{0}^{r_s}\mathrm{d}r\Big( \sqrt{-\bar g}\mathcal{L}\Big)_{BH}\int_{0}^{t_0}dv\nn\\
 &=&\fft{\omega_{n-2}}{16\pi G}\Big[\int_{0}^{r_s}\mathrm{d}r\Big(\sqrt{-\bar g}\mathcal{L}\Big)_{vac}\,\Big(2r_*^{vac}(r_s)-2r_*^{vac}(r) \Big)\nn\\
 &&\qquad\quad-2\int_{r_s}^{\infty}\mathrm{d}r\Big(\sqrt{-\bar g}\mathcal{L}\Big)_{BH}\,r_*^{BH}(r)+t_0\int_{0}^{r_s}\mathrm{d}r\Big( \sqrt{-\bar g}\mathcal{L}\Big)_{BH}\Big]\,.
\eea
It follows that its time derivative is given by
\bea \label{totalbulkactionv2}
\fft{dS_{bulk}}{dt_0}&=&\fft{\omega_{n-2}}{16\pi G}\Big[\int_{0}^{r_s}\mathrm{d}r\Big( \sqrt{-\bar g}\mathcal{L}\Big)_{BH} -\ft{w_{BH}(r_s)f_{BH}(r_s)}{f_{vac}(r_s)}\int_{0}^{r_s}\mathrm{d}r \Big( \sqrt{-\bar g} \mathcal{L}\Big)_{vac}\Big] \nn\\
&=&\fft{\omega_{n-2}}{16\pi G}\Big[-r^{n-2}\,\ft{h'_{BH}}{w_{BH}}\Big|_{0}^{r_s}+2\ell^{-2} r_s^{n-1}\ft{w_{BH}(r_s)f_{BH}(r_s)}{f_{vac}(r_s)}  \Big]\,,
\eea
where in the second line we have adopted the relation $\int \mathrm{d}r\,\sqrt{-\bar g}\,\mathcal{L}=-r^{n-2}h'/w$ for static black holes in Einstein-Scalar gravity\footnote{For general static black holes, one has $\int \mathrm{d}r\,\sqrt{-\bar g}\,\mathcal{L}=\sqrt{-\bar g}\,\mathcal{Q}^{rt}$, where $\mathcal{Q}_{ab}$ is the two-form Wald-Iyer Noether charge \cite{Fan:2018qnt}.}.

Next, we evaluate the GH surface term at the future singularity with total time lapse equal to $t_0$. We find
\be
S_{GH}=-\lim_{r\rightarrow 0}\fft{\omega_{n-2}}{16\pi G}\,r^{n-2}w_{BH}f_{BH}\Big(\ft{h'_{BH}}{h_{BH}}+\ft{2(n-2)}{r}\Big)t_0 \,,\ee
Thus,
\be\label{totalGH} \fft{dS_{GH}}{dt_0}=-\lim_{r\rightarrow 0}\fft{\omega_{n-2}}{16\pi G}\,r^{n-2}w_{BH}f_{BH}\Big(\ft{h'_{BH}}{h_{BH}}+\ft{2(n-2)}{r}\Big) \,.\ee
In addition, for the CA-2 duality, one has
\bea \label{CA2growth}
\fft{dS_{\Lambda}}{dt_0}&=&\fft{\omega_{n-2}}{16\pi G}\Big[-\int_{0}^{r_s}\mathrm{d}r\Big( \sqrt{-\bar g}V_{\Lambda}\Big)_{BH} -\ft{w_{BH}(r_s)f_{BH}(r_s)}{f_{vac}(r_s)}\int_{0}^{r_s}\mathrm{d}r  \sqrt{-\bar g}\,\big( -2\Lambda\big) \Big]\nn\\
&=&\fft{\omega_{n-2}}{16\pi G}\Big[-\int_{0}^{r_s}\mathrm{d}r\Big( \sqrt{-\bar g}V_{\Lambda}\Big)_{BH} -(n-2)\ell^{-2} r_s^{n-1}\ft{w_{BH}(r_s)f_{BH}(r_s)}{f_{vac}(r_s)}\Big]\,,
\eea
whilst for the CV-2 duality
\bea \label{CV2growth}
\fft{dS_{V}}{dt_0}&=&\fft{\omega_{n-2}}{16\pi G}\Big[\int_{0}^{r_s}\mathrm{d}r\, \sqrt{-\bar g_{BH}}\big(-2\Lambda\big) -\ft{w_{BH}(r_s)f_{BH}(r_s)}{f_{vac}(r_s)}\int_{0}^{r_s}\mathrm{d}r  \sqrt{-\bar g_{vac}}\big(-2\Lambda\big)\Big]\\
&=&\fft{\omega_{n-2}}{16\pi G}\Big[(n-1)(n-2)\ell^{-2}\int_{0}^{r_s}\mathrm{d}r\, \sqrt{-\bar g_{BH}} -(n-2)\ell^{-2} r_s^{n-1}\ft{w_{BH}(r_s)f_{BH}(r_s)}{f_{vac}(r_s)}\Big]\,.\nn
\eea
In the following, we will adopt these formulas to numerically study the evolution of the complexity for the various proposals.

\subsubsection{Example 1}

\begin{figure}[htbp]
\centering
\includegraphics[width=140pt]{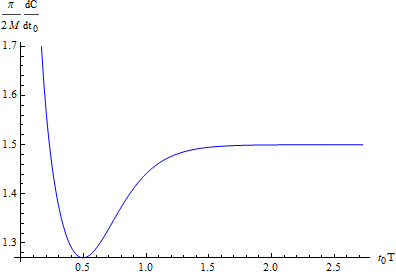}
\includegraphics[width=140pt]{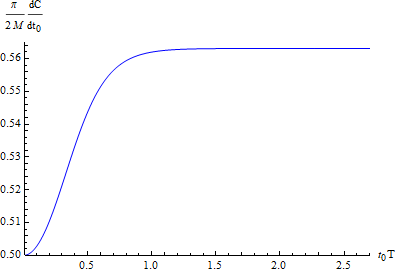}
\includegraphics[width=140pt]{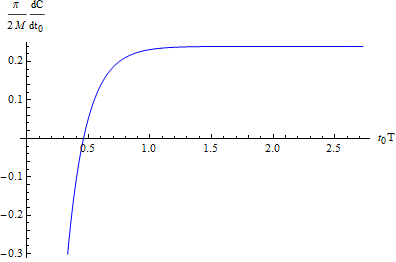}
\caption{{\it The rate of change of complexity for the hairy black hole (\ref{zxf}) with planar topology. From left to right, the various panels correspond to CA/CA-2/CV-2 duality, respectively. We have set $G=\alpha=\ell=q=1\,,\omega_2=4\pi$.}}
\label{zxf0}\end{figure}
\begin{figure}[ht]
\centering
\includegraphics[width=140pt]{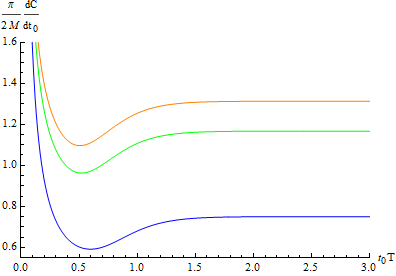}
\includegraphics[width=140pt]{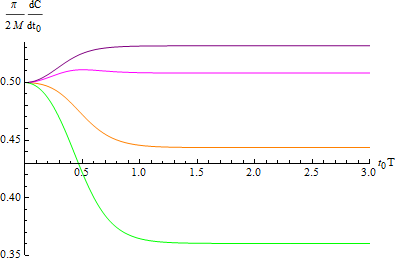}
\includegraphics[width=140pt]{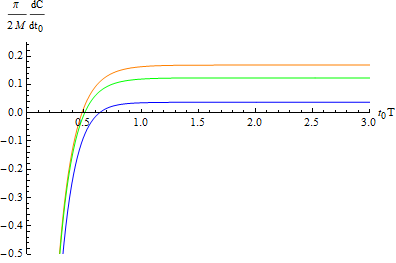}
\caption{{\it Complexity growth of the hairy black hole (\ref{zxf}) with spherical topology. From left to right, the various panels correspond to CA/CA-2/CV-2 duality, respectively. We have $T=0.34$ (Blue), $T=0.46$ (Green), $T=0.61$ (Orange), $T=0.91$ (Magenta), $T=1.21$ (Purple), where $q=2\,,3\,,4\,,6\,,8$ respectively. We have set $G=\alpha=\ell=1\,,\omega_2=4\pi$.}}
\label{zxf1}\end{figure}

First, we study the complexity for the hairy black hole (\ref{zxf}) with planar or spherical topologies. In Fig.\ref{zxf0} and Fig.\ref{zxf1}, we plot the rate of change of complexity (normalized by the Lloyd bound $2M/\pi$) as a function of the boundary time (normalized by the thermal scale $1/T$). For planar black holes, we find that for each of the proposals, the rate of change of complexity as a function of $t_0T$ is identical for different temperatures. This is owing to the scaling symmetry $r\rightarrow \lambda r\,,(t\,,x^i)\rightarrow \lambda^{-1}(t\,,x^i)$ of AdS planar black holes. For all the three proposals, we observe that in in the late time regime, the growth of complexity monotonically increases and approaches the equilibrium value from below. However, at the early time, the time evolution of complexity behaves significantly different for the three proposals.

For the spherical black hole, the scaling symmetry is breaking. However, the complexity shares similar features for the CA and CV-2 duality at different temperatures while for the CA-2 duality, a new property appears at lower temperatures: $\dot{\mathcal{C}}$ monotonically decreases and approaches the late time limit from above.

To understand the evolution of complexity better, we analyze its behavior half analytically at both the early time and the late time.

{\it\textbf{Early time}}

At the early time, the shell radii $r_s$ can be expanded as a Taylor series of the boundary time
\be r_s(t_0)=\fft{2\ell^2}{t_0}-\fft{(q^2\ell^{-2}+8k)}{48}t_0+\fft{\alpha^2q^3}{96\ell^{2}}t_0^2+\cdots  \,,\ee
where the dots denotes higher order terms. Note that the first term is universal, depending only on the asymptotic symmetry of AdS space-time. Evaluating the rate of change of complexity for the three proposals, we obtain
\bea
&&\mathrm{CA}:\quad\qquad  \fft{d\mathcal{C}}{dt_0}=\fft{\omega_2 q^2}{32\pi t_0}+2M-\fft{\omega_2 k q}{8\pi}+\fft{  (q^2+8k\ell^2)M}{32q\alpha^2\ell^4}\,t_0-\fft{23M q^2}{640\ell^4}\,t_0^2+\cdots   \,,\nn\\
&&\mathrm{CA-2}:\quad\, \fft{d\mathcal{C}}{dt_0}=M+\fft{M(q^2-20k\ell^2)}{80\ell^4}\,t_0^2+\cdots   \,,\nn\\
&&\mathrm{CV-2}:\quad\, \fft{d\mathcal{C}}{dt_0}=-\fft{\omega_2 q^2}{8\pi t_0}+\big(1+\ft{3}{2\alpha^2\ell^2}\big)M-\fft{(q^2-4k\ell^2)M}{8q\alpha^2\ell^4 }\,t_0+\cdots   \,.
\eea
It is easily seen that at leading order, the three proposals behaves significantly different: for the CA duality $\dot{\mathcal{C}}$ has a pole with a positive weight while for the CV-2 duality the pole has a negative weight. This implies that once the perturbation is turned on, the complexity increases/decreases logarithmically at the early time for the CA/CV-2 duality while for the CA-2 duality, it grows linearly with the boundary time. This is a characteristic feature, which may serve as an evidence to distinguish the different proposals of complexity.

{\it\textbf{Late time}}

At the late time, we have
\be r_s(t_0)=r_h+c_1 e^{-2\pi T t_0}+\cdots  \,,\ee
where $c_1$ is a positive integration constant. Note that the relaxation time exactly equals to the Lyapunov exponent \cite{Shenker:2013pqa,Roberts:2014isa,Shenker:2014cwa} characterizing the butterfly effect. Substituting the expansion into Eq.(\ref{totalbulkactionv2},\ref{totalGH},\ref{CA2growth},\ref{CV2growth}), we deduce
\bea
&&\mathrm{CA}:\quad\qquad \fft{d\mathcal{C}}{dt_0}=3M-\fft{\omega_2 k q}{8\pi}-\fft{\pi T\mathcal{A}_+ }{r_h}\,c_1 t_0 T e^{-2\pi T t_0}+\cdots   \,,\nn\\
&&\mathrm{CA-2}:\quad\, \fft{d\mathcal{C}}{dt_0}=\fft{\omega_2 r_h^2 \sqrt{4r_h^2+q^2}}{16\pi\ell^2}-\fft{\omega_2\alpha^2 r_h^4}{8\pi(r_h^2+k\ell^2)}\,P_1(z)\,c_1e^{-2\pi T t_0}+\cdots   \,,\nn\\
&&\mathrm{CV-2}:\quad\, \fft{d\mathcal{C}}{dt_0}=\ft{\omega_2\Big(q^3+(2r_h^2-q^2)\sqrt{4r_h^2+q^2} \Big)}{32\pi\ell^2}-\fft{\omega_2\alpha^2 r_h^4}{16\pi(r_h^2+k\ell^2)}\,P_2(z)\,c_1e^{-2\pi T t_0}+\cdots      \,,
\eea
where $\mathcal{A}_+$ denotes the area of the black hole and the functions $P_1(z)\,,P_2(z)$ are defined as ($z\equiv q/r_h$)
\bea
&&P_1(z)=\fft{z^2+6}{\sqrt{z^2+4}}\log{\Big(\fft{\sqrt{z^2+4}+z}{\sqrt{z^2+4}-z} \Big)-3z-\fft{kz^2\sqrt{z^2+4}}{\alpha^2 q^2}}   \,,\nn\\
&&P_2(z)=\fft{12}{\sqrt{z^2+4}}\log{\Big(\fft{\sqrt{z^2+4}+z}{\sqrt{z^2+4}-z} \Big)+z(z^2-6)-\fft{2kz^2\sqrt{z^2+4}}{\alpha^2 q^2}}    \,.
\eea
For the planar black hole $k=0$, the two functions are always positive definite for any given $z>0$. This explains why the growth of complexity in Fig.\ref{zxf0} approaches the late time limit from below. However, for the spherical black hole $k=1$, the situation is not so simple. When $\alpha^2 q^2$ is sufficiently small, the last term in the two functions become important in the small $z$ region. Indeed, we find that $P_1\,,P_2$ become negative in this case. This is why for the CA-2 duality, the complexity growth approaches the late time limit from above at lower temperatures. However, the situation for the CV-2 duality is even more subtle because the existence of the event horizon\footnote{To guarantee the existence of the event horizon, one needs $h(0)=\fft{k}{\alpha^2 q^2}-\fft 12<0$ since $h(r)$ is a monotone increasing function of $r$ and $h(\infty)>0$.} requires $\alpha^2 q^2>2$. In this case, the function $P_2(z)$ is always positive definite so the late time rate of change of complexity is approached from below.

\subsubsection{Example 2}
\begin{figure}[ht]
\centering
\includegraphics[width=140pt]{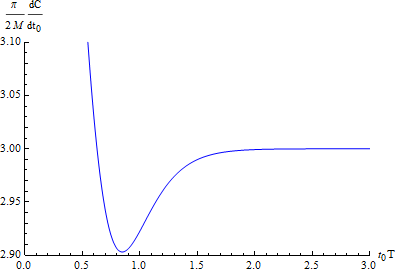}
\includegraphics[width=140pt]{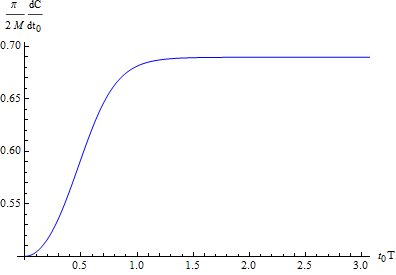}
\includegraphics[width=140pt]{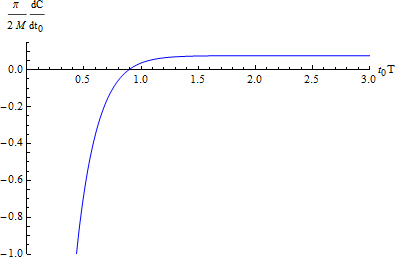}
\caption{{\it The rate of change of complexity for the hairy black hole (\ref{fan}) with $n=4\,,k_0=2\sqrt{2}$. From left to right, the various panels correspond to CA/CA-2/CV-2 duality, respectively. We have set $G=\alpha=\ell=1\,,\omega_2=4\pi$.}}
\label{fanfig}\end{figure}

We continue to study the evolution of complexity for the planar black hole (\ref{fan}). Without loss of generality, we choose a simple case $n=4\,,k_0=2\sqrt{2}$, in which the metric function greatly simplifies to
\be h=r^2\ell^{-2}-3\alpha^2\, q r+3\alpha^2\, r \arctan{\big(\fft{q}{r}\big)} \,.\ee
In fig.\ref{fanfig}, we plot the rate of change of complexity as a function of the boundary time. Again, we observe that the late time limit is approached from below for the three proposals but at the early time the complexity behaves significantly different.

{\it\textbf{Early time}}

At the early time, we have
\be r_s(t_0)=\fft{2\ell^2}{t_0}-\fft{q^2}{6\ell^2}t_0+\fft{\alpha^2q^3}{16\ell^{2}}t_0^2+\cdots  \,.\ee
Evaluating the growth of complexity yields
\bea
&&\mathrm{CA}:\quad\qquad  \fft{d\mathcal{C}}{dt_0}=\fft{\omega_2 q^2}{4\pi t_0}+5M+\fft{ M q}{3\alpha^2\ell^4}\,t_0-\fft{23M q^2}{80\ell^4}\,t_0^2+\cdots   \,,\nn\\
&&\mathrm{CA-2}:\quad\, \fft{d\mathcal{C}}{dt_0}=M+\fft{M q^2}{10\ell^4}\,t_0^2+\cdots   \,,\nn\\
&&\mathrm{CV-2}:\quad\, \fft{d\mathcal{C}}{dt_0}=-\fft{\omega_2 q^2}{\pi t_0}+\big(1+\ft{3\pi}{2\alpha^2\ell^2}\big)M-\fft{5M q}{6\alpha^2\ell^4 }\,t_0+\cdots   \,.
\eea
Again at leading order, we find that the complexity behaves qualitatively similar as the solution (\ref{zxf}) with planar topology.

{\it\textbf{Late time}}

At the late time, we find
\bea
&&\mathrm{CA}:\quad\qquad \fft{d\mathcal{C}}{dt_0}=6M-\fft{6\pi M }{r_h}\,c_1e^{-2\pi T t_0}\,Tt_0+\cdots   \,,\nn\\
&&\mathrm{CA-2}:\quad\, \fft{d\mathcal{C}}{dt_0}=\fft{\omega_2 r_h (r_h^2+q^2) }{8\pi\ell^2}-\fft{3\omega_2\alpha^2 r_h^2}{8\pi}\,\tilde{P}_1(z)\,c_1e^{-2\pi T t_0}+\cdots   \,,\nn\\
&&\mathrm{CV-2}:\quad\, \fft{d\mathcal{C}}{dt_0}=\ft{\omega_2\Big(r_h^3-3q^2r_h+3q^3 \arctan{\big(\fft{r_h}{q} \big)} \Big)}{8\pi\ell^2}-\fft{3\omega_2\alpha^2 r_h^4}{8\pi(r_h^2+q^2)}\,\tilde{P}_2(z)\,c_1e^{-2\pi T t_0}+\cdots      \,,
\eea
where $\tilde{P}_1(z)\,,\tilde{P}_2(z)$ are defined by
\bea
&&\tilde{P}_1(z)=(z^2+3)\arctan{z}-3z   \,,\nn\\
&&\tilde{P}_2(z)=z^3(z^2+1)-3z+3\arctan{z}\,.
\eea
These two functions are always positive definite for $z>0$. Thus, the rate of change of complexity always approaches the late time limit from below for the three proposals.

\subsubsection{Further comment on the early time behaviors}

For the above two solutions, we have found that the complexity has characteristic behaviors at the early time for each of the proposals. As a matter of fact, the early time behavior of complexity is essentially determined by the asymptotic behavior of the black hole solutions. This implies that we may examine it for general static solutions and show to what extent the above features are universal.

Since the large-$r$ expansion of a generally static solution heavily depends on the potential of the scalar field, we first observe that the potential (\ref{potential1}) and (\ref{potential2}) (for $n=4$ dimension) have a same small $\phi$ expansion of the form:
\be\label{smallphiexpansion} V(\phi)=-6g^2-g^2\phi^2+\gamma_4\phi^4+\gamma_5 \phi^5+\cdots \,,\ee
where $\gamma_i$'s are different constants for the two potentials. A standard analysis shows that the asymptotic solution is characterized by three independent parameters, which we take to be $(M\,,\phi_1\,,\phi_2)$
\bea
&&\phi=\fft{\phi_1}{r}+\fft{\phi_2}{r^2}+\cdots\,,\quad w=1-\fft{\phi_1^2}{8r^2}-\fft{\phi_1\phi_2}{3r^3}+\cdots\,,\nn\\
&&f=r^2\ell^{-2}+k+\fft{\phi_1^2}{4\ell^2}-\fft{8\pi M}{\omega_2 r}+\cdots\,.
\eea
Here $(\phi_1\,,\phi_2)$ are two ``charges" of the scalar. However, only two of the parameters are truly independent since $(M\,,\phi_1\,,\phi_2)$ are algebraically constrained by near horizon conditions. Substituting the expansions into Eq.(\ref{tbh2}), we find at the early time
\be r_s=\fft{2\ell^2}{t_0}-\fft{8k\ell^2+\phi_1^2}{48\ell^2}\,t_0+\Big(\fft{\pi  M}{2\omega_2\ell^2}+\fft{\phi_1\phi_2}{48\ell^4} \Big)\,t_0^2+\cdots\,. \ee
It follows that at leading order, the rate of change of complexity behaves as
\bea
&&\mathrm{CA}:\quad\qquad  \fft{d\mathcal{C}}{dt_0}=\fft{\omega_2\phi_1^2}{32\pi t_0}+\mathrm{cons}+\cdots   \,,\nn\\
&&\mathrm{CA-2}:\quad\, \fft{d\mathcal{C}}{dt_0}=\mathrm{cons}+\cdots   \,,\nn\\
&&\mathrm{CV-2}:\quad\, \fft{d\mathcal{C}}{dt_0}=-\fft{\omega_2 \phi_1^2}{8\pi t_0}+\mathrm{cons}+\cdots   \,,
\eea
where the constant terms depend on the full solution. Nevertheless, for generally static solutions in Einstein-Scalar gravity with the potential Eq.(\ref{smallphiexpansion}), the early time behavior of complexity for each of the proposals shares the characteristic features demonstrated previously. Furthermore, these discussions can be straightforward generalized to different type potentials or to higher dimensional solutions, for which we always find similar features for complexity.
%\footnote{In certain cases, for example if the metric functions at asymptotic infinity behave as
%\be f=r^2\ell^{-2}+k+a-\fft{8\pi M}{\omega r}+\cdots\,,\quad w=1-\fft{b_1}{r}-\fft{b_2}{r^2}+\cdots \,,\ee
%the leading order behavior of complexity will be identical for the three proposals, given by $\dot{\mathcal{C}}\propto b_1/t_0^2$.}.
We argue that the early time behavior of holographic complexity following a thermal quench provide a strong evidence to distinguish the various proposals.

\section{Complexity for thermofield double states under a global quench}

In this section, we turn to study the evolution of complexity for eternal black holes perturbed by a global shock wave. In the boundary, it is dual to the thermofield double (TFD) states perturbed by an operator inserted at a time $t=-t_w$
\bea
| TFD  \rangle_{pert}  =  \cOR(-t_w)\, | TFD  \rangle = \UR (t_w)\, \cOR\, U^{\dag}_\text{R} (t_w)\, | TFD  \rangle \, , \label{TFDPertState}
\eea
where without loss of generality we assume the operator $\cOR(-t_w)$ is inserted in the right CFT. A simple discussion in \cite{CMM2} shows that the perturbed TFD state depends only on two combinations of the boundary times, $\tR+t_w$ and $\tL-t_w$. We will recover this result easily on the gravity side, as will be shown later.
\begin{figure}
  \centering
\includegraphics[width=300pt]{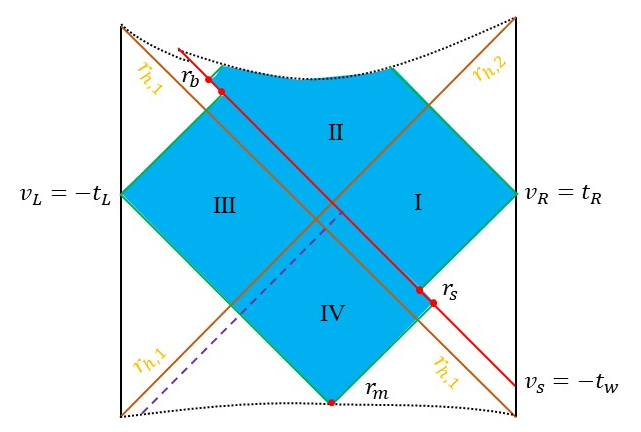}
  	\caption{Penrose-like diagram for one shock wave on an eternal black hole geometry.}
\label{pert}\end{figure}

On the gravity side, the collapse of a thin-shell in an eternal black hole background is shown in the Penrose-like diagram in Fig.\ref{pert}. Just like the onesided black hole case, the spacetime is separated into two static patches by the collapse shell. The information on the time evolution of the spacetime and the complexity is totally contained in the three positions on the WDW patch: $r_b, r_s, r_m$. We identify the three positions as functions of the times $\vL=-\tL$ and $\vR=\tR$ at which the WDW patch is anchored on the left and right boundaries, respectively. From Fig.\ref{pert}, we obtain
\bea\label{tr}
\tL-t_w&=&2\ro(r_b)\,,\nn\\
\tR+t_w&=&-2\rt(r_s)\,,\nn\\
\tL-t_w&=&2\ro(r_s)-2\ro(r_m)\,.
\label{tff}
\eea
Indeed, the time dependence of the evolutions depends only on two combinations of the boundary times, $t_R+t_w$ and $t_L-t_w$, consistent with the analysis in the boundary.

From Eq.(\ref{tff}), one can take the derivative with respect to $t_R$ by fixing $\tL$
\bea\label{dr}
\frac{dr_b}{d\tR}=0\,,\,\,\,\,\frac{dr_s}{d\tR}=-\frac{w_2(r_s)f_{2}(r_s)}{2}, \,\,\,\,\,\,\,\frac{dr_m}{d\tR}=-\frac{w_1(r_m)f_1(r_m)}{2}\frac{w_2(r_s)f_2(r_s)}{w_1(r_s)f_1(r_s)}\,.
\eea
Similarly, when $\tR$ is held fixed, one can deduce the derivative with respect to the left boundary time
\bea\label{dl}
\frac{dr_b}{d\tL}=\frac{w_1(r_b)f_{1}(r_b)}{2}\,,\,\,\,\,\frac{dr_s}{d\tL}=0, \,\,\,\,\,\,\,\frac{dr_m}{d\tL}=-\frac{w_1(r_m)f_1(r_m)}{2}\,.
\eea
From these results, one can study the time evolution for any linear combinations of the two boundary times, namely $\alpha t_R+\beta t_L$. However, in this section we are particularly interested in the time symmetric case $\tL=\tR=t/2$. Hence, we will combine the above results (and the results for the action) in a certain linear combination by using the relation
\be \fft{d}{dt}=\fft 12\Big( \fft{d}{dt_L}+ \fft{d}{dt_R} \Big)\,.\ee

\subsection{Gravitational action and critical times}\label{sec41}
Next we compute the gravitational action on the WDW patch illustrated in Fig.\ref{pert}. The total gravitational action that is relevant to the time evolution can be formally rewritten as
\bea
S_{grav}&=&S_{bulk}+S_{GH}^{(p)}+S_{GH}^{(f)}+\Big[S_{shell}+S_{joint}^{I}(r_b)+S_{joint}^{II}(r_s)\Big]\nn\\
&&+S_{joint}^{III}(r_m)+S_{ct}^{I}(past\,\,right)+S_{ct}^{II}(future\,\,left)+S_{ct}^{III}(past\,\,left)\,,
\eea
where $S_{GH}^{(p)}$/$S_{GH}^{(f)}$ stands for the GH surface term at the past/future singularity, $S_{ct}^{I}(past\,\,right)$ denotes the counterterm action on the past null boundary on the r.h.s of the WDW patch. Here we have dropped a time independent counterterm on the future right null boundary. These various boundary actions are calculated very carefully in Appendix \ref{boundaryaction}. Like the one-sided black hole case, the action on the shell is cancelled by the corner contributions at $r=r_s,r_b$, namely
\be S_{shell}+S_{joint}^{I}(r_b)+S_{joint}^{II}(r_s)=0 \,.\ee

In the following, we focus on dealing with the bulk action. By splitting the WDW patch into several pieces, we deduce
\bea\label{b2}
S_{bulk}&=&\fft{\omega_{n-2}}{16\pi G}\Big\{\int^{r_{\max}}_{r_s}\mathrm{d}r\Big(\sqrt{-\bar g}\mathcal{L}\Big)_{2}\int_{\tR+2\rt(r)}^{\tR}dv+\int^{r_s}_{r_{h,1}}\mathrm{d}r\Big(\sqrt{-\bar g}\mathcal{L}\Big)_{1}\int_{-t_w+2\ro(r)-2\ro(r_s)}^{-t_w}dv \nn\\
&&+\int^{r_{h,1}}_{r_m}\mathrm{d}r\Big(\sqrt{-\bar g}\mathcal{L}\Big)_{1}\int_{-t_w+2\ro(r)-2\ro(r_s)}^{-\tL}dv+\int^{r_{\max}}_{r_{h,1}}\mathrm{d}r\Big(\sqrt{-\bar g}\mathcal{L}\Big)_{1}\int_{-\tL+2\ro(r)}^{-\tL}dv\nn\\
&&+\int^{r_{h,1}}_{r_b}\mathrm{d}r\Big(\sqrt{-\bar g}\mathcal{L}\Big)_{1}\int_{-\tL+2\ro(r)}^{-t_w}dv+\int^{r_b}_{0}\mathrm{d}r\Big(\sqrt{-\bar g}\mathcal{L}\Big)_{2}\int_{-t_w+2\rt(r)-2\rt(r_b)}^{\tR}dv\nn\\
  &&+\int^{r_s}_{r_{b}}\mathrm{d}r\Big(\sqrt{-\bar g}\mathcal{L}\Big)_{2}\int_{-t_{w}}^{\tR}dv\Big\}\nn\\
&=&\fft{\omega_{n-2}}{16\pi G}\Big\{\int^{r_{\max}}_{r_s}\mathrm{d}r\Big(\sqrt{-\bar g}\mathcal{L}\Big)_{2}\Big(-2\rt(r)\Big)+\int^{r_s}_{r_{h,1}}\mathrm{d}r\Big(\sqrt{-\bar g}\mathcal{L}\Big)_{1}\Big(2\ro(r_s)-2\ro(r)\Big) \nn\\
&&+\int^{r_{h,1}}_{r_m}\mathrm{d}r\Big(\sqrt{-\bar g}\mathcal{L}\Big)_{1}\Big(-\tL+t_w-2\ro(r)+2\ro(r_s)\Big)+\int^{r_{\max}}_{r_{h,1}}\mathrm{d}r\Big(\sqrt{-\bar g}\mathcal{L}\Big)_{1}\Big(-2\ro(r)\Big)\nn\\
&&+\int^{r_{h,1}}_{r_b}\mathrm{d}r\Big(\sqrt{-\bar g}\mathcal{L}\Big)_{1}\Big(-t_w+\tL-2\ro(r)\Big)+\int^{r_b}_{0}\mathrm{d}r\Big(\sqrt{-\bar g}\mathcal{L}\Big)_{2}\Big(\tR+t_w-2\rt(r)+2\rt(r_b)\Big)\nn\\
  &&+\int^{r_s}_{r_{b}}\mathrm{d}r\Big(\sqrt{-\bar g}\mathcal{L}\Big)_{2}\big(\tR+t_w\big)\Big\}\,.
\eea
To derive the derivatives of the bulk action with respect to the boundary times, we first calculate some partial derivatives
\bea
&&\fft{\partial S_{bulk}}{\partial r_s}=\fft{\omega_{n-2}}{16\pi G}\fft{2}{w_1(r_s)f_1(r_s)}\int_{r_m}^{r_s}\mathrm{d}r\Big(\sqrt{-\bar g}\mathcal{L}\Big)_1\,,\nn\\
&&\fft{\partial S_{bulk}}{\partial r_b}=\fft{\omega_{n-2}}{16\pi G}\fft{2}{w_2(r_b)f_2(r_b)}\int_{0}^{r_b}\mathrm{d}r\Big(\sqrt{-\bar g}\mathcal{L}\Big)_2\,,\quad \fft{\partial S_{bulk}}{\partial r_m}=0\,,\nn\\
&&\fft{\partial S_{bulk}}{\partial t_R}=\fft{\omega_{n-2}}{16\pi G}\int_0^{r_s}\mathrm{d}r\Big(\sqrt{-\bar g}\mathcal{L}\Big)_2\,,\quad \fft{\partial S_{bulk}}{\partial t_L}=\fft{\omega_{n-2}}{16\pi G}\int_{r_b}^{r_m}\mathrm{d}r\Big(\sqrt{-\bar g}\mathcal{L}\Big)_1\,.
\eea
It follows that the derivatives of the bulk action with respect to the boundary times are given by
\bea\label{br}
\frac{dS_{\bulk}}{d\tR}&=&\fft{\omega_{n-2}}{16\pi G}\Big[-\frac{w_2(r_s)f_2(r_s)}{w_1(r_s)f_1(r_s)}\int^{r_s}_{r_m}dr\Big(\sqrt{-\bar g}\mathcal{L}\Big)_{1}+\int^{r_s}_0dr\Big(\sqrt{-\bar g}\mathcal{L}\Big)_{2}\Big]\,,\nn\\
\frac{dS_{\bulk}}{d\tL}&=&\fft{\omega_{n-2}}{16\pi G}\Big[\frac{w_1(r_b)f_1(r_b)}{w_2(r_b)f_2(r_b)}\int^{r_b}_{0}dr\Big(\sqrt{-\bar g}\mathcal{L}\Big)_{2}+\int^{r_m}_{r_b}dr\Big(\sqrt{-\bar g}\mathcal{L}\Big)_{1}\Big]\,.
\eea
Notice that for Einstein-Scalar gravity, the above bulk integral gives $\int\mathrm{d}r\sqrt{-\bar g}\,\mathcal{L}=-r^{n-2}h'/w$. With respect to the symmetric evolution $\tL=\tR=t/2$, we shall sum the derivatives in a certain linear combination as
\bea
\frac{dS_{\bulk}}{dt}=\frac{1}{2}\Big(\frac{dS_{\bulk}}{d\tR}+\frac{dS_{\bulk}}{d\tL}\Big)\,.
\eea
Note that Eq.(\ref{br}) is valid to the CA-2 and CV-2 duality as well. Explicitly, we have
\bea\label{CA2}
\mathrm{CA-2}:\quad\left\{\begin{array}{ll}
\frac{dS_{\Lambda}}{d\tR}=
\fft{\omega_{n-2}}{16\pi G}\Big[\frac{w_2(r_s)f_2(r_s)}{w_1(r_s)f_1(r_s)}\int^{r_s}_{r_m}dr\big(\sqrt{-\bar g}\,V_\Lambda\big)_{1}-\int^{r_s}_0dr\big(\sqrt{-\bar g}\,V_\Lambda\big)_{2}\Big]\,,\nn\\
\\
\frac{dS_{\Lambda}}{d\tL}=
-\fft{\omega_{n-2}}{16\pi G}\Big[\frac{w_1(r_b)f_1(r_b)}{w_2(r_b)f_2(r_b)}\int^{r_b}_{0}dr\big(\sqrt{-\bar g}\,V_\Lambda\big)_{2}+\int^{r_m}_{r_b}dr\big(\sqrt{-\bar g}\,V_\Lambda\big)_{1}\Big]\,,
\end{array}
\right.
\eea
and
\bea\label{CV2}
\mathrm{CV-2}:\quad\left\{\begin{array}{ll}
\frac{dS_{V}}{d\tR}=
\fft{\Lambda\omega_{n-2}}{8\pi G}\Big[\frac{w_2(r_s)f_2(r_s)}{w_1(r_s)f_1(r_s)}\int^{r_s}_{r_m}dr\big(\sqrt{-\bar g}\,\big)_{1}-\int^{r_s}_0dr\big(\sqrt{-\bar g}\,\big)_{2}\Big]\,,\nn\\
\\
\frac{dS_{V}}{d\tL}=
-\fft{\Lambda\omega_{n-2}}{8\pi G}\Big[\frac{w_1(r_b)f_1(r_b)}{w_2(r_b)f_2(r_b)}\int^{r_b}_{0}dr\big(\sqrt{-\bar g}\,\big)_{2}+\int^{r_m}_{r_b}dr\big(\sqrt{-\bar g}\,\big)_{1}\Big]\,.
\end{array}
\right.
\eea
\\
\textbf{Critical times }\\
There are two important parameters affecting the evolution of complexity: the initial time $-t_w$ where the shock wave was injected into the bulk and the total energy $\Delta M=M_2-M_1$ carried by the shock wave. We will introduce a new parameter $\sigma\equiv T_2/T_1$ to characterize the strength of the shock wave.

 During the collapse, there are several critical times that that are important in describing the dynamical evolution. The first is defined by the time at which $r_m$ lifts off of the past singularity.
 One has
\bea
\tRc1=-t_w-2\rt(r_s)\,,\,\,\,\,\,\,\,\,\,\,\,\,\,\,\,\,\tLc1=t_w+2\ro(r_s)-2\ro(0)\,.
\eea
Similarly, there is a second critical time for the left boundary, which we denote it by $\tLc2$. It is the time at which the crossing point $r_b$ touches the future singularity. One has
\bea
\tLc2=t_w+2\ro(0)\,.
\eea
Since we are interested in the time symmetric case $t_L=t_R=t/2$, we define $\tLc1=\tRc1=t_{c_1}/2$, leading to
\be\label{rsc} t_{c_1}=2t_w-4r_1^*(0)+4r_1^*(r_s)\,,\qquad r_1^*(r_s)+r_2^*(r_s)=-t_w+r_1^*(0) \,,\ee
where the second equation determines the critical value of $r_s$, which will be substituted into the first equation to determine $t_{c_1}$. Likewise, defining $\tLc2=t_{c_2}/2$, we have
\be t_{c_2}=2t_w+4r_1^*(0) \,.\ee
Using these relations, we find
\be t_{c_2}-t_{c_1}=8r_1^*(0)-4r_1^*(r_s) \,.\ee
However, the sign of the difference is undetermined since in our convention, both $r_1^*(0)$ and $r_1^*(r_s)$ are negative. In general, we expect $t_{c_1}<t_{c_2}$ for a sufficiently weak shock wave and $t_{c_1}>t_{c_2}$ for a sufficiently strong shock wave. This is the case that we encounter in sec \ref{hairybtzbh} and sec \ref{higherdim} except for the BTZ black hole which always has $t_{c_1}<t_{c_2}$ due to $r_1^*(0)=0$.

Since there are so many terms that we need to deal with carefully for the CA duality, before doing practical calculations we split these terms into three classes according to the critical times, without considering which one of the two is bigger.\\
$\bullet$ The bulk contribution Eq.(\ref{br}) exists in the full dynamical process. For Einstein-Scalar gravity, we have
\bea\label{bulk}
\frac{dS_{\bulk}}{dt}&=&
\fft{\omega_{n-2}}{32\pi G}\Big\{\fft{w_2(r_s)f_2(r_s)}{w_1(r_s)f_1(r_s)}\,r^{n-2}\fft{h_1'(r)}{w_1(r)}\Big|_{r_m}^{r_s}
-r^{n-2}\fft{h_1'(r)}{w_1(r)}\Big|^{r_m}_{r_b}\\
&&\qquad\qquad-r^{n-2}\fft{h_2'(r)}{w_2(r)}\Big|_{0}^{r_s}
-\frac{w_1(r_b)f_1(r_b)}{w_2(r_b)f_2(r_b)}\,r^{n-2}\fft{h_2'(r)}{w_2(r)}\Big|_{0}^{r_b}\Big\}\nn\\
&=&\fft{\omega_{n-2}}{32\pi G}\Big\{r_s^{n-2}\Big(\fft{w_2(r_s)f_2(r_s)}{w_1(r_s)f_1(r_s)}\fft{h_1'(r_s)}{w_1(r_s)}-\fft{h_2'(r_s)}{w_2(r_s)}\Big)-
\Big(\fft{w_2(r_s)f_2(r_s)}{w_1(r_s)f_1(r_s)}+1\Big)r_m^{n-2}\fft{h_1'(r_m)}{w_1(r_m)} \nn\\
&&\qquad -r_b^{n-2}\Big(\fft{w_1(r_b)f_1(r_b)}{w_2(r_b)f_2(r_b)}\fft{h_2'(r_b)}{w_2(r_b)}-\fft{h_1'(r_b)}{w_1(r_b)}\Big)
+\Big(\fft{w_1(r_b)f_1(r_b)}{w_2(r_b)f_2(r_b)}+1\Big)\epsilon^{n-2}\fft{h_2'(\epsilon)}{w_2(\epsilon)}\Big|_{\epsilon\rightarrow 0}\Big\}\,.\nn
\eea
This term implicitly depends on the critical times $t_{c_1}$ and $t_{c_2}$ through the positions $r_m\,,r_b$. We will take $r_m=0$ for $t<t_{c_1}$ and $r_b=0$ for $t>t_{c_2}$.\\

$\bullet$ When $t<t_{c_1}$, we have the GH surface term Eq.(\ref{GH1}) at the past singularity and the counterterm contributions Eq.(\ref{sct13early}) whilst when $t>t_{c_1}$, we have the corner contributions Eq.(\ref{joint}) and the counterterm contributions Eq.(\ref{sct13late}). We summarize the results in the following\\
\bea
&&t<t_{c_1}\,,\left\{\begin{array}{ll}
\frac{dS_{\GH}^{(p)}}{dt}=\lim_{r\rightarrow 0}\fft{\omega_{n-2}}{32\pi G}\,r^{n-2}w_{1}f_{1}\Big(\ft{h'_{1}}{h_{1}}+\ft{2(n-2)}{r}\Big)
\Big(\frac{w_2(r_s)f_2(r_s)}{w_1(r_s)f_1(r_s)}+1\Big)\,,\\
\\
\fft{d}{dt}\big(S_{ct}^{\oor}+S_{ct}^{\ttr} \big)\Big|_{r_m=0}=\fft{\omega_{n-2}}{32\pi G}r_s^{n-2}
\,w_2(r_s)f_2(r_s)\,
\Big[ \fft{f_2'(r_s)}{f_2(r_s)}+\fft{w_2'(r_s)}{w_2(r_s)}\\
\qquad\qquad\qquad\qquad\qquad\qquad\qquad-\fft{f_1'(r_s)}{f_1(r_s)}-\fft{w_1'(r_s)}{w_1(r_s)}
-\fft{(n-2)}{r_s}\log{\Big( \fft{f_1(r_s)}{f_2(r_s)}\Big)}  \Big]\,,
\end{array}
\right.\label{tc11}\\
&&\nn\\
&&t>t_{c_1}\,,\left\{\begin{array}{ll} \ft{d}{dt}\big(S_{ct}^{\oor}+S_{ct}^{\ttr} \big)=\ft{d}{dt}\big(S_{ct}^{\oor}+S_{ct}^{\ttr} \big)\Big|_{r_m=0} \label{tc12}\\
\qquad\qquad\qquad\quad\,\,-\ft{\omega_{n-2}}{32\pi G}r_m^{n-2}\,w_2(r_s)f_2(r_s)
\Big[ \ft{f_2'(r_s)}{f_2(r_s)}+\ft{w_2'(r_s)}{w_2(r_s)}-\ft{f_1'(r_s)}{f_1(r_s)}-\ft{w_1'(r_s)}{w_1(r_s)}\Big]\\
\qquad\qquad\qquad\quad\,\,+\ft{(n-2)\omega_{n-2}}{16\pi G}r_m^{n-3}\,w_1(r_m)f_1(r_m) \\
\qquad\qquad\qquad\qquad\qquad\quad\Big(\ft{w_2(r_s)f_2(r_s)}{w_1(r_s)f_1(r_s)} +1\Big) \log{\Big( \ft{(n-2)\ell_{ct}\alpha}{w_1(r_m)r_m}\ft{\sqrt{w_1(r_s)f_1(r_s)}}{\sqrt{w_2(r_s)f_2(r_s)}}  \Big)}\,,\\
\\
\frac{dS_{\joint}^{\ttr}}{dt}=\fft{\omega_{n-2}}{32\pi G} r_m^{n-2}\,w_1(r_m)f_1(r_m)\,\\
\qquad\qquad\Big(\fft{w_2(r_s)f_2(r_s)}{w_1(r_s)f_1(r_s)}+1 \Big)\Big[ \fft{h_1'(r_m)}{h_1(r_m)}
+\fft{n-2}{r_m}\log{\Big(\fft{|h_1(r_m)|\,w_2(r_s)f_2(r_s)}{2\alpha^2\,w_1(r_s)f_1(r_s)} \Big)}  \Big]\\
\qquad\qquad+\fft{\omega_{n-2}}{32\pi G} r_m^{n-2}\,w_2(r_s)f_2(r_s)\,\Big(\fft{f_2'(r_s)}{f_2(r_s)}+\fft{w_2'(r_s)}{w_2(r_s)}
-\fft{f_1'(r_s)}{f_1(r_s)}-\fft{w_1'(r_s)}{w_1(r_s)} \Big)  \,.
\end{array}
\right.
\eea
The summation of the Eq.(\ref{tc12}) gives
\bea
&&\frac{d}{dt}\big(S_{ct}^{\oor}+S_{ct}^{\ttr}+S_{\joint}^{\ttr}\big)=\fft{d}{dt}\big(S_{ct}^{\oor}+S_{ct}^{\ttr} \big)\Big|_{r_m=0}\\
&&\qquad\qquad\qquad\qquad\qquad\quad+\fft{\omega_{n-2}}{32\pi G} r_m^{n-2}\,w_1(r_m)f_1(r_m)\Big(\fft{w_2(r_s)f_2(r_s)}{w_1(r_s)f_1(r_s)}+1 \Big)\nn\\
&&\qquad\qquad\qquad\qquad\qquad\qquad\Big[ \fft{h_1'(r_m)}{h_1(r_m)}
+\fft{n-2}{r_m}\log{\Big(\fft{(n-2)^2\ell_{ct}^2\,|f_1(r_m)|}{2r_m^2} \Big)}  \Big]\,,\nn
\eea
which is independent of $\alpha$. In fact, the total gravitational action is independent of the asymptotic normalization of the null norms, as emphasized previously.\\

$\bullet$ When $t<t_{c_2}$, we have the GH surface term Eq.(\ref{GH2}) at the future singularity and the counterterm contributions Eq.(\ref{sct2early}) whilst when $t>t_{c_2}$, we only have Eq.(\ref{GH3}), the GH surface term at the future singularity. We have\\
\bea
&&t<t_{c_2}\,,\left\{\begin{array}{ll}
\frac{dS_{\GH}^{(f)}}{dt}=-\lim_{r\rightarrow 0}\fft{\omega_{n-2}}{32\pi G}\,r^{n-2}w_{2}f_{2}\Big(\ft{h'_{2}}{h_{2}}+\ft{2(n-2)}{r}\Big)
\Big(\frac{w_1(r_b)f_1(r_b)}{w_2(r_b)f_2(r_b)}+1\Big)\,,\\
\\
\fft{dS_{ct}^{\tr}}{dt}=\fft{\omega_{n-2}}{32\pi G}r_b^{n-2}\,w_1(r_b)f_1(r_b)\,\Big[\fft{f_2'(r_b)}{f_2(r_b)}+\fft{w_2'(r_b)}{w_2(r_b)}\\
\qquad\qquad\qquad\qquad\qquad\qquad\quad
-\fft{f_1'(r_b)}{f_1(r_b)}-\fft{w_1'(r_b)}{w_1(r_b)}+\fft{(n-2)}{r_b}\log{\Big( \fft{f_2(r_b)}{f_1(r_b)} \Big)} \Big]\,,
\end{array}
\right.\label{tc21}\\
&&t>t_{c_2}\,,\quad
\ft{dS_{\GH}^{(f)}}{dt}=-\lim_{r\rightarrow 0}\ft{\omega_{n-2}}{32\pi G}\,r^{n-2}\Big(\ft{h'_{2}}{w_{2}}+\ft{h'_1}{w_1}+\ft{2(n-2)}{r}\big(w_2f_2+w_1f_1 \big)\Big)\,.\label{tc22}
\eea

In addition, there is a third critical time at which the WDW patch lifts off of the future singularity. We denote it by $t_{c_0}$. It is determined in a very similar way to finding $t_{c_1}$. We have
\be t_{c_0}=2t_w-4r_2^*(0)+4r_2^*(r_b)\,,\qquad r_1^*(r_b)+r_2^*(r_b)=-t_w+r_2^*(0) \,,\ee
where again the second equation determines the critical value for $r_b$ which should be plugged into the first to find $t_{c_0}$.
In fact, in analogy with the evolution of complexity for eternal black holes \cite{Carmi:2017jqz}, roughly at the time $t=-t_{c_0}$ the growth of complexity enters into a plateau at the early time. We will come to this point in sec \ref{earlylate}. Notice that $t_{c_0}<2t_w$ which is reminiscent of the fact that the shock wave is injected into the bulk at $t=-2t_w$.

With the three critical times, there are three regimes in the evolution, which should be considered separately. For $t_{c_2}>t_{c_1}$, we work in the scenario of
\bea&&\mathrm{scenario\,\,A}\left\{\begin{array}{ll}
\mathrm{I}:\qquad -t_{c_0}<t<t_{c_1}\,,\qquad r_b\,,r_s\neq 0\,,\,r_m=0\,,\qquad \mathrm{Eq}.(\ref{bulk}\,,\ref{tc11}\,,\ref{tc21})\,,\nn\\
\mathrm{II}:\qquad t_{c_1}<t<t_{c_2}\,,\qquad\quad\, r_b\,,r_s\,,r_m\neq 0\,,\qquad\quad \,\mathrm{Eq}.(\ref{bulk}\,,\ref{tc12}\,,\ref{tc21})\,,\nn\\
\mathrm{III}:\qquad\quad t>t_{c_2}\,,\qquad\quad\, r_s\,,r_m\neq 0\,,\,r_b=0\,,\qquad \,\,\mathrm{Eq}.(\ref{bulk}\,,\ref{tc12}\,,\ref{tc22})\,.\nn
\end{array}
\right.
\eea
In the first regime I, the total action includes the bulk contributions Eq.(\ref{bulk}) with $r_m=0$ as well as Eq.(\ref{tc11}) and Eq.(\ref{tc21}). For the second regime II, we need Eq.(\ref{bulk}), Eq.(\ref{tc12}) and Eq.(\ref{tc21}) whilst for the third regime, we need Eq.(\ref{bulk}) with $r_b=0$, Eq.(\ref{tc12}) and Eq.(\ref{tc22}).

 When $t_{c_1}>t_{c_2}$, we will work in a different scenario. The three regimes that need to be considered are:
\bea&&\mathrm{scenario\,\,A'}\left\{\begin{array}{ll}
\mathrm{I}':\qquad -t_{c_0}<t<t_{c_2}\,,\qquad r_b\,,r_s\neq 0\,,\,r_m=0\,,\qquad\,\, \mathrm{Eq}.(\ref{bulk}\,,\ref{tc11}\,,\ref{tc21})\,,\nn\\
\mathrm{II}':\qquad t_{c_2}<t<t_{c_1}\,,\qquad\, r_s\neq 0\,,r_b=0=r_m\,,\qquad \mathrm{Eq}.(\ref{bulk}\,,\ref{tc11}\,,\ref{tc22})\,,\nn\\
\mathrm{III}':\qquad\quad t>t_{c_1}\,,\qquad\quad\, r_s\,,r_m\neq 0\,,\,r_b=0\,,\qquad \,\,\,\,\mathrm{Eq}.(\ref{bulk}\,,\ref{tc12}\,,\ref{tc22})\,.\nn
\end{array}
\right.
\eea
The difference from the scenario A is in the second regime II', we deal with equations Eq.(\ref{tc11})\,,Eq.(\ref{tc22}), instead of Eq.(\ref{tc12})\,,Eq.(\ref{tc21}) in the regime II. It is worth emphasizing that in both cases, we deal with the same set of equations at the early time and the late time.

For the CA-2 and CV-2 duality, we just have a single bulk term Eq.(\ref{CA2}). Thus, we have
\bea\label{CAV2}
&&\frac{dS_{\Lambda}}{dt}=\fft{\omega_{n-2}}{32\pi G}\Big[\frac{w_2(r_s)f_2(r_s)}{w_1(r_s)f_1(r_s)}\int^{r_s}_{r_m}dr\big(\sqrt{-\bar g}\,V_\Lambda\big)_{1}-\int^{r_m}_{r_b}dr\big(\sqrt{-\bar g}\,V_\Lambda\big)_{1}\nn\\
&&\qquad\qquad\qquad\qquad-\int^{r_s}_0dr\big(\sqrt{-\bar g}\,V_\Lambda\big)_{2}-\frac{w_1(r_b)f_1(r_b)}{w_2(r_b)f_2(r_b)}\int^{r_b}_{0}dr\big(\sqrt{-\bar g}\,V_\Lambda\big)_{2}\Big]  \,,\nn\\
&&\frac{dS_{V}}{dt}=
\fft{\Lambda\omega_{n-2}}{16\pi G}\Big[\frac{w_2(r_s)f_2(r_s)}{w_1(r_s)f_1(r_s)}\int^{r_s}_{r_m}dr\big(\sqrt{-\bar g}\,\big)_{1}-\int^{r_m}_{r_b}dr\big(\sqrt{-\bar g}\,\big)_{1}\nn\\
&&\qquad\qquad\qquad\qquad-\int^{r_s}_0dr\big(\sqrt{-\bar g}\,\big)_{2}-\frac{w_1(r_b)f_1(r_b)}{w_2(r_b)f_2(r_b)}\int^{r_b}_{0}dr\big(\sqrt{-\bar g}\,\big)_{2}\Big]   \,.
\eea
Now we are ready to perform numerical calculations for the critical times $t_{c_0}\,,t_{c_1}\,,t_{c_2}$, the evolution of the positions $r_s\,,r_b\,,r_m$ as well as the evolution of complexity.

\subsection{Universal features at the early and late time}\label{earlylate}
There are two simple limits for the rate of change of complexity. The first is at the early time $t_w\rightarrow \infty$, one has $r_s\rightarrow r_{h_2}\,,r_b\rightarrow r_{h_1}\,,r_m=0$. It is straightforward to derive the rate of change of complexity in this limit. For the CA duality, we find
\bea \fft{d\mathcal{C}}{dt}\Big|_{t_w\rightarrow \infty}
&=&\fft12\big(\dot{\mathcal{C}}_{2\,,\mathrm{late}}-\dot{\mathcal{C}}_{1\,,\mathrm{late}} \big)
+O\Big(T_1(2t_w-t)e^{-\pi T_1(2t_w-t)}\Big)\,,
\eea
where $\mathcal{C}_1\,,\mathcal{C}_2$ denote the complexity of the initial and final eternal black holes which have metric components $(h_1\,,f_1\,,w_1)$ and $(h_2\,,f_2\,,w_2)$, respectively. We have
\bea
&&\mathrm{CA}:\quad \dot{\mathcal{C}}_{\mathrm{late}}=-\lim_{\epsilon\rightarrow 0} \fft{(n-2)\omega_{n-2}}{8\pi G}\epsilon^{n-3}w_1(\epsilon)f_1(\epsilon)\,,\nn\\
&&\mathrm{CA}-2:\quad\dot{\mathcal{C}}_{\mathrm{late}}=-\fft{\omega_{n-2}}{16\pi G}\int^{r_{h}}_{0}dr \sqrt{-\bar g}\,V_\Lambda \,,\nn\\
&&\mathrm{CV}-2:\quad\dot{\mathcal{C}}_{\mathrm{late}}=
-\fft{\Lambda\omega_{n-2}}{8\pi G}\int^{r_{h}}_{0}dr \sqrt{-\bar g}\,.
\eea
For the CA-2 and CV-2 duality, we find
\bea \fft{d\mathcal{C}}{dt}\Big|_{t_w\rightarrow \infty}
&=&\fft12\big(\dot{\mathcal{C}}_{2\,,\mathrm{late}}-\dot{\mathcal{C}}_{1\,,\mathrm{late}} \big)
+O\big(e^{-\pi T_1(2t_w-t)}\big)\,.
\eea
 It is easily seen that for each of the proposals in this limit, the leading order for the rate of change of complexity is given by the average of the difference between the growth rate of the initial and final eternal black holes. In fact, there may exist a certain regime $-t_{c_0}<t<t_{c_{1\,,2}}$ at the early time in which the rate of change of complexity is approximately a constant, given by the above early time limit at leading order. This happens when $r_s$ has already approached its equilibrium value exponentially fast in the time regime close to the critical time $t_{c_{1\,,2}}$ while $r_b$ still decays exponentially slow, namely
\bea\label{earlytimerb}
&&r_s=r_{h_2}\Big(1+c_s\, e^{-\pi T_2(t+2t_w)}+\cdots \Big)\,,\nn\\
&&r_b=r_{h_1}\Big(1-c_b\, e^{-\pi T_1(2t_w-t)}+\cdots \Big)  \,,
\eea
where $c_s\,,c_b$ are positive constants. This leads to a first plateau in the rate of change of complexity. We may define the ``boundary" of the plateau as
\bea
&&r_s=r_{h_2}\Big(1+c_s\, e^{-\gamma}+\cdots \Big)\,,\nn\\
&&r_b=r_{h_1}\Big(1-c_b\, e^{-\gamma}+\cdots \Big)  \,,
\eea
where $\gamma>0$ is a constant of order 1. Then the condition on the shock wave parameter is
\be t_w>\gamma(\sigma+1)/4\pi T_2\,,\ee
where $\sigma\equiv T_2/T_1$ characterizes the strength of the injected shock wave. We will work in this case in the remaining sections, where we show that there indeed exists a plateau at the early time for the rate of change of complexity for the three proposals. The region of the plateau is roughly given by
\be \gamma/\pi T_2-2t_w \lesssim t\lesssim 2t_w-\gamma/\pi T_1\,.\ee
Furthermore, when $t_{c_1}>t_{c_2}$, a second plateau will emerge in the regime $t_{c_2}<t<t_{c_1}$. This happens if $t_{c_2}>\gamma/\pi T_2-2t_w$ since in this case $r_b=r_m=0$ while $r_s$ exponentially approaches the equilibrium value.

Another interesting limit is at the late time $t\rightarrow \infty$, $r_s\rightarrow r_{h_2}\,,r_m\rightarrow r_{h_1}$. We have
\bea\label{latetimersrm}
r_m=r_{h_1}\Big(1-c_m\, e^{-\pi T_1(t-2t_w)}+\cdots \Big)\,,
\eea
where $c_m$ is a positive constant with dimension 1. The late time behavior of the position $r_s$ was already specified in Eq.(\ref{earlytimerb}). We will show that the next-to-leading order term of $r_m$ essentially determines the behavior of complexity at the late time. To leading order, we find
\bea\label{bulkformation0}
&&\fft{dS}{dt}\Big|_{t\rightarrow \infty}=-\lim_{\epsilon\rightarrow 0} \fft{(n-2)\omega_{n-2}}{16\pi G}\epsilon^{n-3}\Big(w_1(\epsilon)f_1(\epsilon)+w_2(\epsilon)f_2(\epsilon)\Big)+\cdots\,,\nn\\
&&\frac{dS_{\Lambda}}{dt}\Big|_{t\rightarrow \infty}=-\fft{\omega_{n-2}}{32\pi G}\Big[\int^{r_{h_1}}_{0}dr\big(\sqrt{-\bar g}\,V_\Lambda\big)_{1}+\int^{r_{h_2}}_0dr\big(\sqrt{-\bar g}\,V_\Lambda\big)_{2}\Big]+\cdots  \,,\nn\\
&&\frac{dS_{V}}{dt}\Big|_{t\rightarrow \infty}=
-\fft{\Lambda\omega_{n-2}}{16\pi G}\Big[\int^{r_{h_1}}_{0}dr\big(\sqrt{-\bar g}\,\big)_{1}+\int^{r_{h_2}}_0dr\big(\sqrt{-\bar g}\,\big)_{2}\Big]  +\cdots \,.
\eea
Thus, for the three proposals, we have at leading order
\be \dot{\mathcal{C}}_{\mathrm{late}}=\fft12\big(\dot{\mathcal{C}}_{1\,,\mathrm{late}}+\dot{\mathcal{C}}_{2\,,\mathrm{late}}\big)
+\cdots\,. \ee
However, the next-to-leading order term is important since its sign determines how the late time rate of change of complexity is approached. For the CA duality, we find that this term is proportional to $T_1 t\,e^{-\pi T_1(t-2t_w)}$ which is positive definite
\be \dot{\mathcal{C}}_{\mathrm{late}}=\fft12\big(\dot{\mathcal{C}}_{1\,,\mathrm{late}}+\dot{\mathcal{C}}_{2\,,\mathrm{late}}\big)+
\fft{1}{2}(n-2)T_1 S_1\,c_m  T_1(t-2t_w)e^{-\pi T_1(t-2t_w)}+\cdots \,.\ee
Thus, for the CA duality the late time limit is always approached from above. For the CA-2 and CV-2 duality, the next-to-leading order term is proportional to $e^{-\pi T_1(t-2t_w)}$ which however has a negative sign
\be\dot{\mathcal{C}}_{\mathrm{late}}=\fft12\big(\dot{\mathcal{C}}_{1\,,\mathrm{late}}+\dot{\mathcal{C}}_{2\,,\mathrm{late}}\big)
-\fft{\omega_{n-2}}{32\pi^2 G}\,c_m r_{h_1}^{n-1} w_1(r_{h_1})|V(r_{h_1})|\,e^{-\pi T_1(t-2t_w)}+\cdots \,,\ee
where $V=V_\Lambda$ for the CA-2 duality and $V=2\Lambda$ for the CV-2 duality respectively. Hence, the late time limit is always approached from below.

We will show these universal features for the evolution of complexity for certain hairy black hole solutions.

\subsection{Hairy BTZ black hole}\label{hairybtzbh}

There is a particularly simple solution contained in (\ref{fan}). In $n=3$ dimension, by taking $\alpha^2=\ft{1}{128}(8-k_0^2)(16-k_0^2)$, we obtain the simplified scalar potential
\be\label{scalarpotential} V(\phi)=-\ft{1}{8}g^2\big(16+k_0^2\sinh^2{\Phi} \big)\big(\cosh{\Phi}\big)^{\ft{k_0^2}{4}} \,.\ee
For later purpose, we focus on $0<k_0<2\sqrt{2}$ such that the resulting solution has and only has one event horizon.
The corresponding static solution reads
\bea\label{hairybtz}
&&\phi(r)=k_0\,\mathrm{arcsinh}\Big(\sqrt{\fft{\lambda r_h}{r}}\, \Big)\,,\qquad w(r)=\Big(1+\fft{\lambda r_h}{r} \Big)^{(1-\lambda)/\lambda}\,,\nn\\
&&h(r)=g^2(r-r_h)(r+\lambda r_h)\,w(r)\,,\quad f(r)=g^2(r-r_h)(r+\lambda r_h)\,w(r)^{-1}\,,
\eea
where we have introduced a new parameter
\be \lambda=\fft{8}{8-k_0^2}\,,\qquad \mathrm{or}\qquad k_0=2\sqrt{\fft{2(\lambda-1)}{\lambda}} \,.\ee
Note that $\lambda\geq 1$. The equality is taken when $k_0=0$, where the solution reduces to the BTZ black hole.
It is easily seen that the above solution is a one-parameter generalization of the BTZ black hole. We would like to call it {\it hairy BTZ black hole}. The temperature and the black hole mass are given by
\be T=\fft{(\lambda+1)r_h}{4\pi\ell^2}\,,\quad M=\fft{(\lambda+1)r_h^2}{16 G\ell^2} \,.\ee
\begin{figure}[ht]
\centering
\includegraphics[width=200pt]{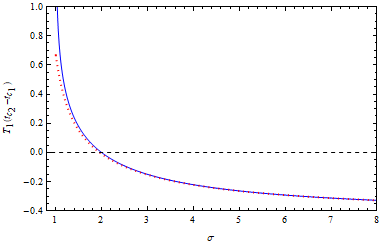}
\includegraphics[width=190pt]{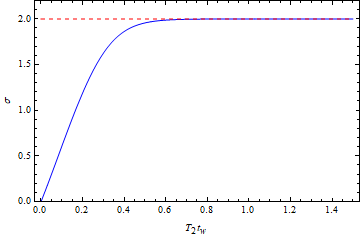}
\caption{{\it In the left panel, the difference of the critical times $T_1(t_{c_2}-t_{c_1})$ is plotted as a function of the shock wave parameter $\sigma$ for $T_2t_w=1.5$ (solid) and $T_2t_w=0.5$ (dashed). The difference is a monotone decreasing function of $\sigma$ and becomes negative for sufficiently large $\sigma$. In the right panel, the critical $\sigma$ at which $t_{c_2}=t_{c_1}$ is plotted as a function of $t_w$. It monotonically increases and tends to 2 in the large $t_w$ limit. Here we have set $\lambda=2$.}}
\label{critictime}\end{figure}
Using the shock wave parameter $\sigma\equiv T_2/T_1$, we have
\be \fft{M_2}{M_1}=\sigma^2\,,\qquad \fft{S_2}{S_1}=\sigma \,.\ee
The tortoise coordinate can be solved as
\be r^*(r)=\ft{\ell^2}{(\lambda+1)r_h}\log{\Big(\fft{|r-r_h|}{r+\lambda r_h}\Big)} \,.\ee
Notice that $r^*(0)=0$ when $\lambda=1$. In this case, one always has $t_{c_2}>t_{c_1}$ for the BTZ black hole. However, for $\lambda>1$, $r^*(0)$ is no longer zero. This significantly affects the critical times $t_{c_1}\,,t_{c_2}$. In Fig.\ref{critictime}, we plot the difference of the critical times $T_1(t_{c_2}-t_{c_1})$ as a function of the shock wave parameter $\sigma$. We observe that for any given $t_w$, the difference monotonically decreases as $\sigma$ increases and approaches a negative value for sufficiently large $\sigma$. The critical $\sigma_c$ at which the two critical times are equal to one another is determined by
\be r_s=\fft{\lambda\,r_{h_1}}{\lambda-1}\quad \Rightarrow\quad 4\pi T_2t_w=\big( \sigma_c+1\big)\log\lambda-\log{\Big(\fft{\lambda-(\lambda-1)\sigma_c}{1+(\lambda-1)\sigma_c} \Big)} \,.\ee
It turns out that the functional relation between the critical $\sigma_c$ and $t_w$ depends on $\lambda$. Nonetheless, we find that for sufficiently large $t_w$, $\sigma_c$ monotonically increases and approaches a constant $\lambda/(\lambda-1)$ in the large $t_w$ limit, as shown in the right panel of Fig.\ref{critictime}. Thus, for $\sigma<\sigma_c$ we will work in the scenario A and for $\sigma>\sigma_c$ we will work in the scenario A' instead.

\begin{figure}[htbp]
\centering
\includegraphics[width=200pt]{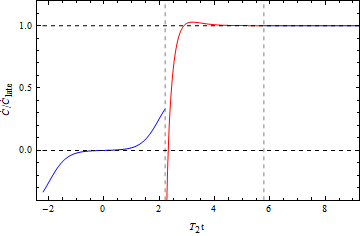}
\includegraphics[width=200pt]{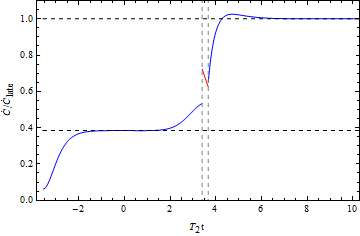}
\includegraphics[width=200pt]{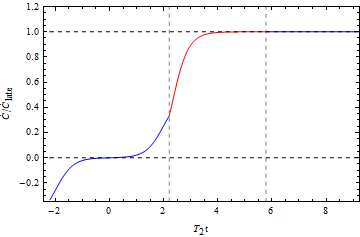}
\includegraphics[width=200pt]{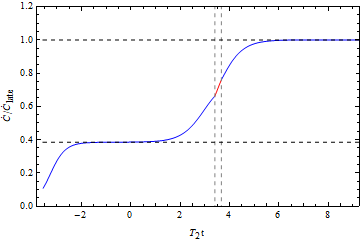}
\includegraphics[width=200pt]{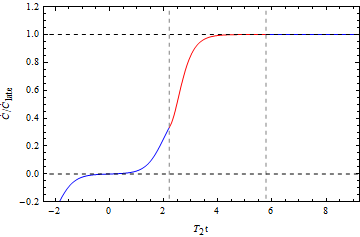}
\includegraphics[width=200pt]{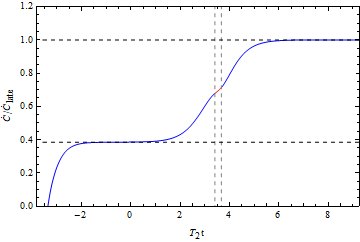}
\caption{{\it The evolution of complexity for hairy BTZ black hole perturbed by either a weak shock wave $\sigma=1+10^{-5}$ with $T_2t_w=3$ (left panels) or a strong shock wave $\sigma=3/2$ with $T_2t_w=2$ (right panels). In both cases, $t_{c_1}<t_{c_2}$. The panels from top to bottom correspond to CA/CA-2/CV-2 duality, respectively. In all these panels, the vertical dashed lines correspond to $t_{c_1}$ and $t_{c_2}$ respectively whilst the horizontal dashed lines correspond to the plateau $\dot{\mathcal{C}}/{\dot{\mathcal{C}}_{\mathrm{late}}}=(M_2-M_1)/(M_2+M_1)$ and the late time limit respectively.}}
\label{hairy1}\end{figure}
From Eq.(\ref{tff}), the positions $r_s\,,r_b\,,r_m$ can be solved exactly as
\bea\label{hbtzrsrbrm}
&&r_s=r_{h_2}\,\fft{1+\lambda e^{-2\pi T_2(t_R+t_w)}}{1-e^{-2\pi T_2(t_R+t_w)}}\,,\nn\\
&&r_b=r_{h_1}\,\fft{e^{-2\pi T_1(t_L-t_w)}-\lambda}{e^{-2\pi T_1(t_L-t_w)}+1}\,,\nn\\
&&r_m=r_{h_1}\,\fft{r_s+\lambda r_{h_1}-\lambda(r_s-r_{h_1})e^{-2\pi T_1(t_L-t_w)}}{r_s+\lambda r_{h_1}+(r_s-r_{h_1})e^{-2\pi T_1(t_L-t_w)}}\,,
\eea
where again when $\lambda=1$, the results reduce to the BTZ black hole case \cite{CMM2}.  With these relations in hand, we are ready to numerically calculate the rate of change of complexity for the various proposals. Without loss of generality, we will set $\lambda=2=k_0$ in the numerical calculations.

We first consider the $t_{c_1}<t_{c_2}$ case. In the left panels of Fig.\ref{hairy1}, we show the time evolution of complexity for a weak shock wave with $\sigma=1+10^{-5}$ and $T_2t_w=3$. We find that for the CA duality there exists a negative spike at $t=t_{c_1}$ and the growth of complexity continuously crosses the critical point $t=t_{c_2}$ and then approaches the late time limit from above. However, for the CA-2/CV-2 duality, the growth of complexity smoothly increases\footnote{However, generally the second derivative of the complexity with respect to the boundary time are discontinuous at the critical times for both weak and strong shock waves.} and approaches the late time limit from below. A common feature shared by the three proposals is there exists a plateau at the early time $-t_{c_0}\lesssim t\lesssim t_{c_1}$ in which the rate of change of complexity remains a constant approximately.

In the right panels of Fig.\ref{hairy1}, we study the evolution of complexity for a strong shock wave with $\sigma=3/2$ and $T_2t_w=2$. For the CA duality, there is a positive spike at $t=t_{c_1}$ but the rate of change of complexity jumps at the second critical time $t=t_{c_2}$ as well. On the contrary, for the CA-2/CV-2 duality, the rate of change of complexity still smoothly increases without any discontinuity at the critical times.

\begin{figure}[htbp]
\centering
\includegraphics[width=230pt]{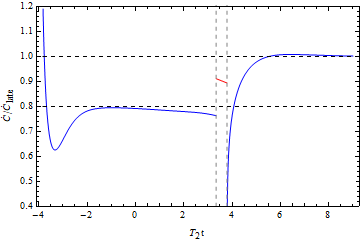}
\includegraphics[width=210pt]{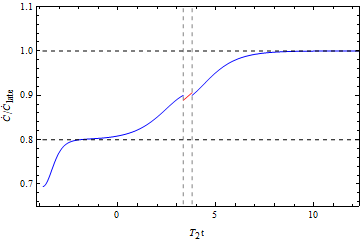}
\includegraphics[width=210pt]{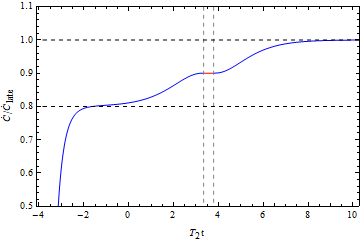}
\caption{{\it The evolution of complexity for hairy BTZ black hole perturbed by a strong shock wave $\sigma=3$ with $T_2t_w=2$ which leads to $t_{c_2}<t_{c_1}$. The vertical dashed lines correspond to $t_{c_2}\,,t_{c_1}$ respectively and the horizontal dashed lines correspond to the plateau $\dot{\mathcal{C}}/{\dot{\mathcal{C}}_{\mathrm{late}}}=(M_2-M_1)/(M_2+M_1)$ and the late time limit respectively. The three panels correspond to CA (top), CA-2 (bottom left) and CV-2 (bottom right) duality, respectively.}}
\label{hairy2}\end{figure}

 When the shock wave is sufficiently strong, we will instead have $t_{c_1}>t_{c_2}$. In Fig.\ref{hairy2}, we study the evolution of complexity for $\sigma=3\,,T_2t_w=2$. We observe that for all the three proposals, the width of the plateau at the early time is strongly suppressed and there is a transit region $t_{c_2}<t<t_{c_1}$, where the rate of change of complexity remains approximately a constant. In addition, for the CA/CA-2 duality the growth of complexity is discontinuous at the critical times, though the weights of the spikes are different.

\subsection{Higher dimensional hairy black hole}\label{higherdim}
We continue to study a $n=4$ dimensional black hole included in (\ref{fan}). For $k_0=2\sqrt{5}\,,\alpha^2=2\sqrt{2}g^2$, the scalar potential reads
\be V(\phi)=-2g^2\cosh^5{\big( \ft{\phi}{2\sqrt{5}}\big)}\Big(5\cosh^3{\big( \ft{\phi}{2\sqrt{5}}\big)}-2\cosh^5{\big( \ft{\phi}{2\sqrt{5}}\big)}+4\sqrt{2}\sinh^5{\big( \ft{\phi}{2\sqrt{5}}\big)} \Big) \,.\ee
The solution greatly simplifies to
\be \phi=2\sqrt{5}\,\mathrm{arcsinh}\big(\fft{r_h}{r}\big)\,,\quad w=\ft{r^5}{\big(r^2+r_h^2 \big)^{5/2}}\,,\quad
h=g^2r^2\Big(1-\ft{2\sqrt{2}r_h^3}{\big( r^2+r_h^2\big)^{3/2}} \Big) \,,\ee
so we can analytically integrate the tortoise coordinate. Outside the black hole event horizon, we have
\bea\label{hdtortoise}
r^*(r)&=&\fft{\ell^2}{6\sqrt{2}r_h}\Big[- i\pi+\gamma+3\log{\big(\varphi(r)+1 \big)}\\
&&\qquad\qquad+\log{\big(\varphi(r)-3-2\sqrt{2}\,\big)}-a\log{\big(\varphi(r)-b \big)}-a^*\log{\big(\varphi(r)-b^*\big)} \Big] \,,\nn
\eea
where the $-i\pi$ factor in the square bracket should be dropped in the black hole interior. The function $\varphi(r)$ is defined as
\be \varphi(r)=\Big( \fft{r_h}{r}+\sqrt{1+\big(\fft{r_h}{r} \big)^2}\, \Big)^2 \,.\ee
The constants $a,b,\gamma$ are given by
\bea
&&\gamma=a\log{\big(1-b \big)}+a^*\log{\big(1-b^* \big)}-4\log{2}-\log{\big(1+\sqrt{2} \big)}\,,\nn\\
&&a=2+\sqrt{3}\,i\,,\qquad b=\fft17\big(3-\sqrt{2} \big)\big(1-\sqrt{6} i\big)\,.
\eea
Note that $\gamma$ is real and $\gamma<0$. In the central of spacetime, one has
\be r^*(0)=\fft{\gamma\, \ell^2}{6\sqrt{2}r_h} \,,\ee
which is nonzero.
\begin{figure}[htbp]
\centering
\includegraphics[width=200pt]{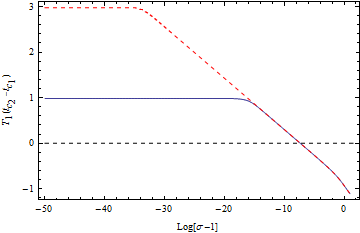}
\includegraphics[width=200pt]{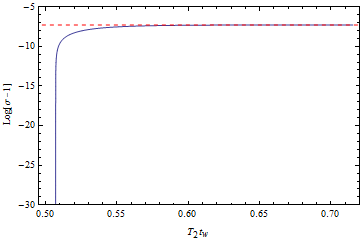}
\caption{{\it In the left panel, the difference between the critical times $T_1(t_{c_2}-t_{c_1})$ is plotted as a function of $\log{(\sigma-1)}$ for $T_2t_w=1$ (solid) and $T_2t_w=2$ (dashed), respectively. In both cases, the critical shock wave parameter is given by $\sigma_c\simeq 1.00066882$. In fact, this is the upper bound on the critical shock wave parameter, which can be seen from the right panel. $\sigma_c$ is a monotone increasing function of $t_w$ and approaches the upper bound in the large $t_w$ limit. }}
\label{hdtimediff}\end{figure}
Thus, the relation between the two critical times $t_{c_1}$ and $t_{c_2}$ depends heavily on the shock wave parameters $(t_w\,,\sigma)$. We find that the difference $t_{c_2}-t_{c_1}$ is positive for a sufficiently weak shock wave and becomes negative for a stronger shock wave. For example, for a weak shock wave with $\sigma=1+10^{-4}\,,t_w=2/T_2$, we have $t_{c_1}=3.11000187/T_2<t_{c_2}= 3.32392288/T_2$ whilst for a strong shock wave with $\sigma=2\,,t_w=2/T_2$, we have $t_{c_1}=4.50000151/T_2>t_{c_2}= 2.64798097/T_2$. In Fig.\ref{hdtimediff}, we show that for fixed $t_w$, the difference between the critical times is a monotone decreasing function of $\sigma$. It is positive when $\sigma<\sigma_c$ and becomes negative when $\sigma>\sigma_c$. The critical point is determined by
\bea
i\pi+\gamma&=&3\log{\big(\varphi_1(r_s)+1 \big)}+\log{\big(\varphi_1(r_s)-3-2\sqrt{2}\,\big)}\nn\\
&&\quad-a\log{\big(\varphi_1(r_s)-b \big)}-a^*\log{\big(\varphi_1(r_s)-b^*\big)}
\,,\eea
which gives rise to the critical radii $r_s\simeq 1.00066856\,r_{h_1}$. This constrains the shock wave parameters $(t_w\,,\sigma)$ through Eq.(\ref{rsc}). We find
\bea
4\pi T_2t_w&=&-\fft{1}{2\sqrt{2}}\Big(-i\pi+(\sigma_c+1)\,\gamma+3\log{\big(\varphi_2(r_s)+1 \big)}+\log{\big(\varphi_2(r_s)-3-2\sqrt{2}\,\big)} \nn\\
&&\qquad\qquad-a\log{\big(\varphi_2(r_s)-b \big)}-a^*\log{\big(\varphi_2(r_s)-b^*\big)}\Big)\,.
\eea
By numerically solving this equation, we find that $\sigma_c$ is a monotone increasing function of $t_w$ and approaches an upper bound, given by $\sigma_c\simeq 1.00066882$ in the large $t_w$ limit. This is shown in the right panel of Fig.\ref{hdtimediff}.

\begin{figure}[htbp]
\centering
\includegraphics[width=205pt]{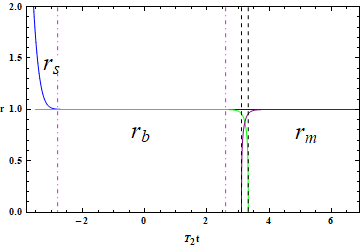}
\includegraphics[width=200pt]{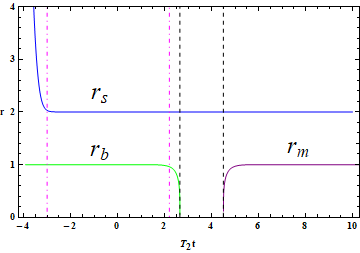}
\caption{{\it The time evolution of $r_s$ (blue), $r_b$ (green) and $r_m$ (purple) for the four dimensional hairy black hole perturbed by either a weak shock wave $\sigma=1+10^{-4}$ which has $t_{c_1}<t_{c_2}$ or a strong shock wave $\sigma=2$ which $t_{c_1}>t_{c_2}$. In both cases, we set $T_2t_w=2$. The vertical dashed lines correspond to $t_{c_1}\,,t_{c_2}$ respectively. Roughly, the dotdashed lines correspond to the first plateau of the growth of complexity at initial times.}}
\label{rsrbrm}\end{figure}

Given the tortoise coordinate Eq.(\ref{hdtortoise}), the time evolution of the three positions $r_s\,,r_b\,,r_m$ have already been solved analytically as implicit functions of the boundary time in Eq.(\ref{tff}). Since their explicit expressions are lacking, to gain a physical intuition on their behaviors, we show them as functions of the boundary time in Fig.\ref{rsrbrm}.

\begin{figure}[htbp]
\centering
\includegraphics[width=200pt]{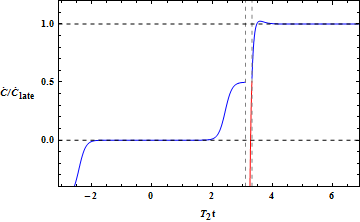}
\includegraphics[width=200pt]{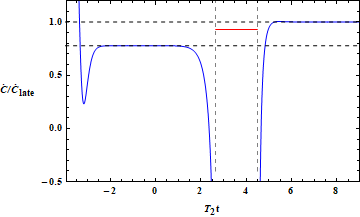}
\includegraphics[width=200pt]{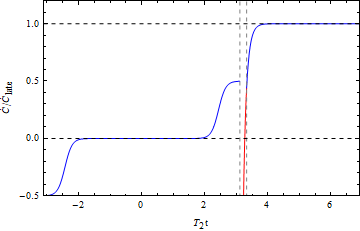}
\includegraphics[width=200pt]{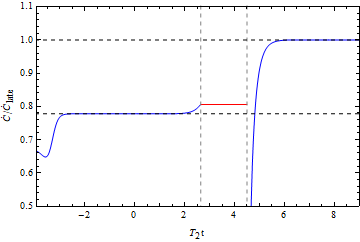}
\includegraphics[width=200pt]{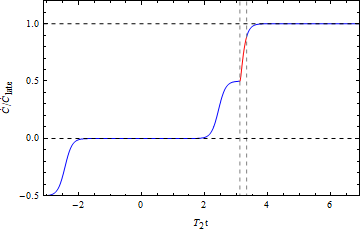}
\includegraphics[width=200pt]{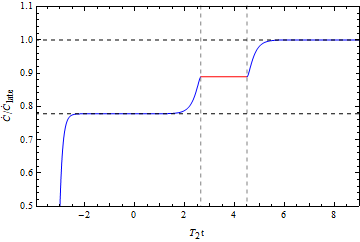}
\caption{{\it The time evolution of complexity for a higher dimensional hairy black hole perturbed by either a weak shock wave $\sigma=1+10^{-4}$ with $t_{c_1}<t_{c_2}$ (left panels) or a strong shock wave $\sigma=2$ with $t_{c_1}>t_{c_2}$ (right panels). In both cases, we have set $T_2t_w=2$ . The panels from top to bottom correspond to CA/CA-2/CV-2 duality, respectively. In all these panels, the vertical dashed lines correspond to $t_{c_1}$ and $t_{c_2}$ respectively whilst horizontal dashed lines correspond to the plateau $\dot{\mathcal{C}}/{\dot{\mathcal{C}}_{\mathrm{late}}}=(M_2-M_1)/(M_2+M_1)$ and the late time limit respectively.}}
\label{hdhairy}\end{figure}

In the left panel, we have $t_{c_1}<t_{c_2}$ and in the right panel, we have $t_{c_1}>t_{c_2}$. In both cases, we find that at the early time $-t_{c_0}<t<t_{c_1}$, there exists an overlap regime, where both $r_s\,,r_b$ exponentially approach their equilibrium value, namely in this regime $r_s\simeq r_{h_2}\,,r_b\simeq r_{h_1}$. This corresponds to the first plateau for the rate of change of complexity, as shown in the left panel Fig.\ref{hdhairy}, where we study the evolution of complexity for a weak shock wave $\sigma=1+10^{-4}$ with $T_2t_w=2$. We are aware of that there exists a negative spike at $t=t_{c_1}$ for both CA and CA-2 duality.

In the right panel of Fig.\ref{rsrbrm}, there is a second plateau existing in the transit regime $t_{c_2}<t<t_{c_1}$ where $r_s\simeq r_{h_2}\,,r_b=r_m=0$. This leads to a second plateau for the rate of change of complexity, as shown in the right panel of Fig.\ref{hdhairy}. There we study the evolution of complexity for a strong shock wave $\sigma=2$ with $T_2t_w=2$. Again, we observe that at $t=t_{c_1}$, there is a negative spike for both CA and CA-2 duality. However, for the CA duality there is a positive spike at $t=t_{c_2}$ whilst for the CV-2 duality, the rate of change of complexity are always continuous at the critical times.

We conclude that these different features in the quench process can be used to distinguish the three proposals of holographic complexity.

\subsection{Complexity of formation}

In the following, we turn to evaluate the complexity of formation for the perturbed state dual to the shock wave geometry. We follow the procedure established in \cite{Chapman:2016hwi}. The complexity of formation is defined by subtracting the contributions of two copies of AdS vacuum from the complexity of the geometries of interest. For example, for static black holes, one has
\be \Delta \mathcal{C}\equiv \mathcal{C}(\mathrm{BH})-2\,\mathcal{C}(\mathrm{AdS\,vacua}) \,.\ee
It was shown in \cite{Chapman:2016hwi} that for the CA and CV duality, this quantity is manifestly finite with all the divergences removed by the vacuum contributions. However, for the CA-2 and CV-2 proposal, the complexity of formation may have logarithmic divergences for certain black holes, which can not be removed by vacuum contributions\footnote{In the presence of a scalar field, the UV structure of complexity may be changed since new divergent terms which are absent in the vacuum, may emerge for certain cases.}.

By taking $\tL=\tR=0$ and $r_m=0$, the calculations of gravitational actions in the section \ref{sec41} can be applied in this situation as well. First, from Eq.(\ref{tff}) we have following identities
\bea\label{rsrbtw}
-t_w&=&2\ro(r_b)=2\rt(r_s)\,.
\eea
However, notice that the third equality in Eq.(\ref{tff}) does not hold any longer since the position $r_m$ is immersed in the past singularity.

From the bulk integral Eq.(\ref{b2}), we deduce
\bea\label{bulkformation}
\Delta S_{bulk}
&=&\fft{\omega_{n-2}}{16\pi G}\Big[\int^{r_{\max}}_{r_s}\mathrm{d}r\Big(\sqrt{-\bar g}\mathcal{L}\Big)_{2}\Big(-2\rt(r)\Big)+\int^{r_s}_{0}\mathrm{d}r\Big(\sqrt{-\bar g}\mathcal{L}\Big)_{1}\Big(t_w+2\ro(r_s)-2\ro(r)\Big) \nn\\
&&+\int^{r_b}_{0}\mathrm{d}r\Big(\sqrt{-\bar g}\mathcal{L}\Big)_{2}\Big(t_w-2\rt(r)+2\rt(r_b)\Big)+\int^{r_{\max}}_{r_b}\mathrm{d}r\Big(\sqrt{-\bar g}\mathcal{L}\Big)_{1}\Big(-2\ro(r)\Big)\nn\\
&&+\int^{r_s}_{r_{b}}\mathrm{d}r\Big[\Big(\sqrt{-\bar g}\mathcal{L}\Big)_{2}-\Big(\sqrt{-\bar g}\mathcal{L}\Big)_{1}\Big]t_w\Big]-2S_{bulk,vac}\nn\\
&=&\fft{\omega_{n-2}}{8\pi G}\Big[-I_0+\int^{r_b}_{0}\mathrm{d}r \beta_2(r)+\int^{r_{max}}_{r_s}\mathrm{d}r \beta_2(r)\nn\\
&&\qquad\qquad\qquad\qquad+\int^{r_s}_{0}\mathrm{d}r \beta_1(r)+\int^{r_{max}}_{r_b}\mathrm{d}r \beta_1(r)
\Big]-2S_{bulk,vac}\,,
\eea
where we have introduced
\bea I_0&=&\alpha_1(r_{max})r_1^*(r_{max})+\alpha_2(r_{max})r_2^*(r_{max}) \nn\\
&&+\Big(\ft{1}{2}t_w-r_1^*(0)+r_1^*(r_s) \Big)\alpha_1(0)+\Big(\ft{1}{2}t_w-r_2^*(0)+r_2^*(r_b) \Big)\alpha_1(0)\,,
\eea
and
\bea
\alpha(r)=-r^{n-2}\frac{h'(r)}{w(r)}\,,\quad \beta(r)&=&-r^{n-2}\frac{h'(r)}{h(r)}\,.
\eea
For the boundary contributions, from Eq.(\ref{GHP}) and Eq.(\ref{GHF}), we have
\bea\label{GHtw}
\Delta S_{\GH}^{(p)}&=&-\lim_{r\rightarrow 0}\fft{\omega_{n-2}}{16\pi G}\,r^{n-2}w_{1}f_{1}\Big(\ft{h'_{1}}{h_{1}}+\ft{2(n-2)}{r}\Big)\big(t_w+2\ro(r_s)-2\ro(0)\big)\,,\nn\\
\Delta S_{\GH}^{(f)}&=&-\lim_{r\rightarrow 0}\fft{\omega_{n-2}}{16\pi G}\,r^{n-2}w_{2}f_{2}\Big(\ft{h'_{2}}{h_{2}}+\ft{2(n-2)}{r}\Big)\big(t_w+2\rt(r_b)-2\rt(0)\big) \,.
\eea
Next, we consider the counterterms in the equations (\ref{sct10}), (\ref{sct2}), (\ref{sct3}) and (\ref{ctf}). We have
\bea
\Delta S_{ct}^{\oor}
&=&\fft{\omega_{n-2}}{8\pi G}\Big[ r_s^{n-2}\log{\Big(\fft{\tilde\alpha}{\alpha}\Big)}+\int^{r_s}_{0}\mathrm{d}\big( r^{n-2}\big)\log{\Big(\fft{w_2}{w_1}\Big)}-\int_{0}^{r_{max}}\mathrm{d}\big( r^{n-2}\big)\log{w_2}\Big]\,,\nn\\
\Delta S_{ct}^{\tr}
&=&\fft{\omega_{n-2}}{8\pi G}\Big[ r_b^{n-2}\log{\Big(\fft{\hat\alpha}{\alpha}\Big)}+\int^{r_b}_0\mathrm{d}\big( r^{n-2}\big)\log{\Big(\fft{w_1}{w_2}\Big)}
-\int^{r_{max}}_{0}\mathrm{d}\big( r^{n-2}\big)\log{w_1}\Big]\,,\nn\\
\Delta S_{ct}^{\ttr}&=&-\fft{\omega_{n-2}}{8\pi G}\int^{r_{max}}_{0}\mathrm{d}\big( r^{n-2}\big)\log{w_1}\,,\nn\\
\Delta S_{ct}^{\fr}&=&-\fft{\omega_{n-2}}{8\pi G}\int^{r_{max}}_{0}\mathrm{d}\big( r^{n-2}\big)\log{w_2}\,.
\eea
Thus we have
\bea
\Delta S_{ct}&=&\fft{\omega_{n-2}}{8\pi G}\Big[ r_s^{n-2}\log{\Big(\fft{\tilde\alpha}{\alpha}\Big)}+r_b^{n-2}\log{\Big(\fft{\hat\alpha}{\alpha}\Big)}\nn\\
&&\qquad\quad+\int^{r_s}_{r_b}\mathrm{d}\big( r^{n-2}\big)\log{\Big(\fft{w_2}{w_1}\Big)}\Big]-2\int_{0}^{r_{max}}\mathrm{d}\big( r^{n-2}\big)\log{\big(w_1 w_2\big)}\Big]\,.
\eea
It turns out that the joint contributions at the positions $r_s\,,r_b\,,r_m$ are trivial: $\Delta S_{\joint}^{r_s\,,r_b\,,r_m}=0$ according to Eq.(\ref{jointrsrb}) and Eq.(\ref{jointiii}).

In addition, to remove all the divergences we need consider the contributions of surface terms and joint terms at the UV cut off. One finds
\bea
\Delta S_{GH}^{cut}&=&-\fft{\omega_{n-2}}{8\pi G}\Big(p_2(r_{max})\rt(r_{max})+p_1(r_{max})\ro(r_{max})\Big)\,,\nn\\
\Delta S_{\joint}^{cut}&=&\fft{\omega_{n-2}}{8\pi G}r_{max}^{n-2}\Big(\log h_2(r_{max})+\log h_1(r_{max})\Big)\,,
\eea
where
\bea
p(r)\equiv r^{n-2}w(r)f(r)\Big(\frac{h^\prime(r)}{h(r)}+\frac{2(n-2)}{r}\Big)\,.
\eea
Thus for the CA duality, the desired complexity of formation can be obtained as follows
\bea
\Delta \mathcal{C}=\fft{\Delta S_{bulk}+\Delta S_{GH}+\Delta S_{ct}+\Delta S_{joint}}{\pi}\,.
\eea
For the CA-2 and CV-2 proposal, one only has the bulk term with $\mathcal{L}=-V_{\Lambda}$ for the former and $\mathcal{L}=-2\Lambda$ for the latter, respectively.

Before moving to numerical calculations, we shall comment on the complexity of formation in the large $t_w$ limit, namely $t_w\rightarrow \infty$ where one has $r_s\rightarrow r_{h_2}\,,r_b\rightarrow r_{h_1}$. Taking the derivative of (\ref{bulkformation}) with respect to $t_w$, one finds
\be\label{bulkderivative} \fft{d\Delta S_{bulk}}{dt_w}=\fft{\omega_{n-2}}{16\pi G}\Big[\int_0^{r_s}\mathrm{d}r\,\Big(\sqrt{-\bar g}\,\mathcal{L} \Big)_2+\int_0^{r_b}\mathrm{d}r\,\Big(\sqrt{-\bar g}\,\mathcal{L} \Big)_1 \Big]\,.\ee
Hence, for the CA-2 and CV-2 proposal, one has
\be k_\infty\equiv\fft{d\Delta \mathcal{C}}{dt_w}\Big|_{t_w\rightarrow \infty}=2\fft{d\mathcal{C}}{dt}\Big|_{t\rightarrow \infty} \,,\ee
as can be easily seen from Eq.(\ref{bulkformation0}). In fact, this identity is valid for the CA duality as well because of
\be \fft{d}{dt_w}\big(\Delta S_{GH}^{(p)}+\Delta S_{GH}^{(f)} \big)\Big|_{t_w\rightarrow \infty}=2\fft{dS_{GH}}{dt}\Big|_{t\rightarrow \infty} \,,\ee
according to Eq.(\ref{GHtw}) and Eq.(\ref{GH3}) and
\be \fft{dr_{s,b}}{dt_w}=2\fft{dr_{s,b}}{dt}\quad\Rightarrow\quad \fft{d \Delta S_{ct}}{dt_w}\Big|_{t_w\rightarrow \infty}=2\fft{dS_{ct}}{dt}\Big|_{t\rightarrow \infty} \,,\ee
according to the equations (\ref{dr}), (\ref{dl}) and (\ref{rsrbtw}), respectively.
\begin{figure}
  \centering
  \includegraphics[width=140pt]{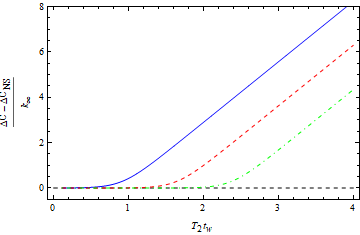}
  \includegraphics[width=140pt]{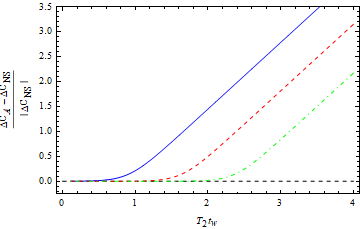}
  \includegraphics[width=140pt]{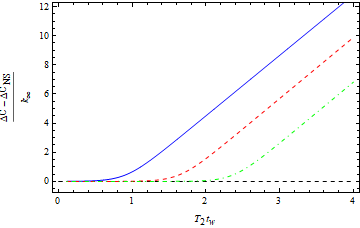}
  \caption{Complexity of formation for hairy BTZ black hole with $\lambda=2$ perturbed by a weak shock wave. The three panels from left to right correspond to CA/CA-2/CV-2 duality, respectively. In all the panels, $\sigma=1+10^{-2}$ (blue solid), $\sigma=1+10^{-4}$ (red dashed) and $\sigma=1+10^{-6}$ (green dotdashed). For all the three proposals, the complexity of formation roughly grows linearly with $t_w$ after a scrambling time $t_{scr}=\fft{1}{2\pi T_1}\log{\big(3/\epsilon \big)}$. }
\label{formationcomplexity}\end{figure}

Thus, for the three proposals, the formation of complexity grows linearly with $t_w$ in the large $t_w$ regime, with the slope given by the double of the late time rate of change of complexity in the full time evolution. From these results, we naturally expect the switchback effect exists for all the three proposals.

In the following, we will calculate the complexity of formation for a certain black hole: the Hairy BTZ black hole with $\lambda=2$. In this case, we have
\bea
&&\mathrm{CA}:\qquad\quad k_\infty=\fft{8}{3\pi}\big(M_1+M_2\big)\,,\nn\\
&&\mathrm{CA}-2:\quad \,k_\infty=\fft{4}{3\pi}\big(M_1+M_2\big)\,,\nn\\
&&\mathrm{CV}-2:\quad \,k_\infty=\fft{2}{\pi}\Big(\log{\big(2+\sqrt{3}\,\big)}-\ft{2}{\sqrt{3}} \Big)\big(M_1+M_2\big)\,.
\eea
For the unperturbed black hole, we obtain by subtracting the complexity of the Neveu-Schwarz vacuum ($k=1$)
\bea
&&\mathrm{CA}:\qquad\quad \Delta \mathcal{C}_{NS}=-\frac{\ell}{2G}+\ft{3+6\log{\big(\fft{\ell}{r_h}\big)}-5\log{2}}{3\pi^2}\,S\,,\nn\\
&&\mathrm{CA}-2:\quad \,\Delta \mathcal{C}_{NS}=\fft{\pi\ell}{4G}+\ft{4\log2-9}{6\pi^2}\,S\,,\nn\\
&&\mathrm{CV}-2:\quad \,\Delta \mathcal{C}_{NS}=\fft{3S}{\pi}\log{\Big(\fft{\delta}{\ell} \Big)}+\mathrm{const}\,,
\eea
where $\delta$ is the UV cut off in the Fefferman-Graham coordinate. We find that for the CA and CA-2 duality, the ratio of complexity of formation to the black hole entropy (or the entanglement entropy of the TFD state) is a constant in the high temperature limit. However, for the CV-2 duality the complexity of formation is logarithmically divergent. Its coefficient contains universal information: the entanglement entropy about the dual state.

For the shock wave geometries, we would like to compare the complexity of formation with the result of the unperturbed black hole. The numerical results for the three proposals are shown in the panels of Fig.\ref{formationcomplexity} respectively. It is clear that the switchback effect exists for all the three proposals. To understand this better, we shall analyze it semi-analytically as follows.

For a very weak shock wave $\sigma=1+\epsilon$ at the regime $T_2t_w\gg1$, one finds from Eq.(\ref{hbtzrsrbrm})
\bea
x_s\equiv\frac{r_s}{r_{h_2}}=1+\eta+\mathcal{O}(\epsilon\eta,\eta^2)\,,\qquad x_b\equiv\frac{r_b}{r_{h_1}}=1-\eta+\mathcal{O}(\eta^2)\,.
\eea
where $\eta\equiv3e^{-2\pi T_1t_w}$. In this limit, there are two interesting regimes to consider: $\epsilon\ll \eta$ and $\epsilon\gg\eta$. The scrambling time $t_{scr}=\frac{1}{2\pi T_1}\log{\big(3/\epsilon\big)}$ is determined by the transition condition $\epsilon=\eta$. In the first regime, one finds $\Delta \mathcal{C}-\Delta \mathcal{C}_{NS}\simeq 0$ owing to $\sigma\simeq 1$ whilst in the second regime, one finds instead
\be \Delta \mathcal{C}-\Delta \mathcal{C}_{NS}=k_\infty(t_w-t_{scr})+\mathcal{O}(\epsilon) \,.\ee
For example, for the CA duality one has
\bea
\Delta \mathcal{C}-\Delta \mathcal{C}_{NS}&=&\frac{r_{h_1}}{12\pi G}\Big[3\log\Big(\ft{x_s-1}{\sigma-x_b}\Big)+\log\Big(\ft{\sigma x_s-1}{1-x_b}\Big)+4\log\Big(\ft{\sigma-x_b}{x_s-1}\Big)\nn\\
&&\qquad\quad+3\log\Big(\ft{3(\sigma x_s-1)}{x_s^2+x_s-2}\Big)+3\log\Big(\ft{3(\sigma-x_b)}{2-x_b-x_b^2}\Big)\Big] +\mathcal{O}(\epsilon,\eta)\nn\\
&=&\frac{2r_{h_1}}{3\pi G}\Big[2\pi T_1t_w+\log(\epsilon/3)\Big]+\mathcal{O}(\epsilon)\nn\\
&=&\frac{16M_1}{3\pi}(t_w-t_{scr})+\mathcal{O}(\epsilon)\,,
\eea
and for the CA-2 duality
\bea
\Delta \mathcal{C}-\Delta \mathcal{C}_{NS}&=&\frac{r_{h_1}}{6\pi G}\log\Big(\ft{(\sigma x_s-1)(w-x_b)}{(1-x_b)(x_s-1)}\Big)+\mathcal{O}(\epsilon,\eta)\nn\\
&=&\frac{r_{h_1}}{3\pi G}\Big[2\pi T_1t_w+\log(\epsilon/3)\Big]+\mathcal{O}(\epsilon)\nn\\
&=&\frac{8M_1}{3\pi}(t_w-t_{scr})+\mathcal{O}(\epsilon)\,.
\eea
For the CV-2 duality, we arrive at the similar result but the intermediate expressions are lengthy which are not instructive to present. Hence, for all the three proposals we can approximate the complexity of formation in both regimes with the following simple expression:
\bea
\Delta \mathcal{C}\simeq\Delta \mathcal{C}_{NS}+\Theta(t_w-t_{scr})\,k_\infty(t_w-t_{scr})\,.
\eea
It is worth emphasizing that the scrambling time is a physical parameter that is independent of the holographic proposals of complexity. The differences for the various proposals are containded in the complexity of formation for the unperturbed black hole and the rate of growth after the scrambling time.

\section{Discussions}
In this paper, we studied the evolution of complexity following a global quantum quench for various holographic proposals, which relate the complexity to certain gravitational objects defined on the Wheeler-DeWitt patch. We focus on two dynamical hairy black holes (\ref{zxf}) and (\ref{fan}). In addition to the known ``Complexity=Action" (CA) duality, we studied the ``Complexity=Volume 2.0" (CV-2) duality and ``Complexity=Action 2.0" (CA-2) duality. The former postulates that the complexity is dual to the spacetime volume of the WDW patch whilst the latter relates the complexity to the non-derivative gravitational action relevant to the cosmological constant on the WDW patch.

We find that surprisingly, all these different proposals reproduce some known properties of complexity, such as the linear growth rate at late times and the switchback effect for TFD states perturbed by global shock waves. However, each of these proposals also has its own characteristic features during the dynamical evolution. We briefly summarize our main results as follows.

First, for both of the hairy black holes (\ref{zxf}) and (\ref{fan}), we find that after a thermal quench the complexity at early times logarithmically increases (decreaes) for the CA (CV-2) duality while for the CA-2 duality, it grows linearly with the boundary time. This is a characteristic feature for the three proposals and it is essentially determined by the mass square of the scalar field with $m^2=-2\ell^{-2}$ (in other words it does not depend on the detail of the black hole model). Of course, for a different mass square, the leading order behavior of complexity will be changed but in general it is significantly different for different proposals. Hence, we argue that the early time behavior of complexity following a thermal quench may serve as a powerful tool to distinguish the various proposals of complexity.

 We further studied the evolution of complexity for TFD states perturbed by a global shock wave. In this case, the behavior of complexity highly depends on two critical times $t_{c_1}$ and $t_{c_2}$ ( note that we have chosen the boundary times $t_L=t_R$). By studying the two hairy black holes, we find that the rate of change of complexity behaves significantly different at the critical times for the various proposals. However, they also share some universal features. For example, at early times the complexity growth rate is given by the average of the difference between the growth rate of the initial and final eternal black holes, namely
 \bea \fft{d\mathcal{C}}{dt}\Big|_{t_w\rightarrow \infty}
&=&\fft12\big(\dot{\mathcal{C}}_{2\,,\mathrm{late}}-\dot{\mathcal{C}}_{1\,,\mathrm{late}} \big)\,,
\eea
while at late times it is given by
\be \fft{d\mathcal{C}}{dt}\Big|_{\mathrm{late}}=\fft12\big(\dot{\mathcal{C}}_{2\,,\mathrm{late}}+\dot{\mathcal{C}}_{1\,,\mathrm{late}}\big)
+\cdots\,. \ee
In particular, the next-to-leading order at late times is always negative definite and hence the complexity growth rate generally approaches the late time limit from below.

Moreover, we compute the complexity of formation for shock wave geometries for the three proposals and find that they all show the switchback effect
\bea
\Delta \mathcal{C}\simeq\Delta \mathcal{C}_{NS}+\Theta(t_w-t_{scr})\,k_\infty(t_w-t_{scr})\,,
\eea
where $\Delta \mathcal{C}_{NS}$ is the complexity of formation for eternal black holes (without shock waves), $t_{scr}$ is the scrambling time and $k_\infty=\dot{\mathcal{C}}_{2\,,\mathrm{late}}+\dot{\mathcal{C}}_{1\,,\mathrm{late}}$ is the double of the complexity growth rate at late times for the shock wave geometries. 

It is worth emphasizing that the above features are universal in the sense that they hold for all the three proposals and they do not depend on the detail of the black hole model. Notice that while we have found in some circumstance the three conjectures for complexity can be distinguished, we still can not claim which of the conjecture is better than the other according to the results in this paper. To achieve this goal, deeper studies are needed. 

There are a list of interesting directions that deserve further investigations.

$\bullet$ Our results show that the complexity proposed by the CA-2 and CV-2 duality contains similar data as that of the CA duality. However, they are computationally much simpler than the latter. It is interesting to search whether there are more such gravitational objects measuring the complexity of the boundary state. In particular, does there exist an object that has only boundary terms on the WDW patch?

$\bullet$Recently, the action principle and its growth rate for higher derivative gravities with nonsmooth boundaries was well studied in \cite{Cano:2018ckq,Cano:2018aqi,Jiang:2018sqj,Jiang:2018pfk}. Based on the results in these papers, it is interesting to further investigate the evolution of complexity following a local or global quench for general higher derivative gravities.

$\bullet$ Though we focus on the complexity dual to the total system on the boundary, the holographic subregion complexity is also an interesting subject which has attracted a lot of attentions in recent years (see for example \cite{Alishahiha:2015rta,Ben-Ami:2016qex,Carmi:2016wjl,Abt:2017pmf,Agon:2018zso,Abt:2018ywl,Chen:2018mcc}). In fact, there are also several different proposals in literature for the subregion complexity. It is interesting to investigate how these proposals behave following a global quantum quench and whether they contain similar data or have characteristic features during the dynamical evolution.

$\bullet$ Recently, the complexity for free field theories has been widely examined. However, it is difficult to compare the results there with those in holography. Indeed, there is not a priori reason to expect the results on both sides should agree since holographic theories are strongly coupled, with a large number of degrees of freedoms. In order to compare the results on both sides directly, it is of great importance to study the complexity for field theories in the large N limit. A recent paper \cite{Caputa:2018kdj} may provide a guidance towards this direction.

\section*{Acknowledgments}
We are grateful to Chuyu Lai, Hugo Marrochio and Robert C. Myers for useful comments and discussions. Z.Y. Fan is supported in part by the National Natural Science Foundations of China (NNSFC) with Grant No. 11805041, No. 11873025 and No. 11575270. M. Guo is supported in part by NNSFC Grants No.11775022 and No.11375026 and also supported by the China Scholarship Council. M. Guo also thanks the Perimeter Institute \lq\lq{}Visiting Graduate Fellows\rq\rq{} program. Research at Perimeter Institute is supported in part by the Government of Canada through the Department of Innovation, Science and Economic Development and by the Province of Ontario through the Ministry of Research, Innovation and Science.\\

\appendix
\section{Exact dynamical hairy BTZ black hole}

The Einstein-Scalar gravity with the potential (\ref{scalarpotential}) admits an exact dynamical solution in three dimension, given by
\bea\label{dyfan} ds^2&=&-h\,dv^2+2w\, dr dv+r^2 dx^2\,,\qquad \phi=k_0\,\mathrm{arcsinh}\Big(\sqrt{\fft{\lambda a}{r}}\, \Big)\,,\\
      w&=&\Big(1+\fft{\lambda a}{r} \Big)^{(1-\lambda)/\lambda}\,,\qquad h=\Big( g^2(r-a)(r+\lambda a)-2\lambda \dot a\Big) w\,, \nn\eea
where $a=a(v)$ and a dot denotes the derivative with respect to $v$. In the static limit, the solution is a one-parameter generalization of the BTZ black hole an hence we call it {\it hairy BTZ black hole}. Substituting the solution into equations of motion, one finds  the only non-trivial equation is
\be E_{vv}=\fft{1}{2r a}\Big(2\lambda a\ddot a-2(\lambda-1)\dot{a}^2+(\lambda+1) g^2 a^2\dot a     \Big) \,.\ee
The equation can be integrated, giving rise to
\be 2\dot a+g^2 a^2\Big( 1-\big(\fft{r_{h_2}}{a}\big)^{\fft{\lambda+1}{\lambda}} \Big)=0 \,.\ee
Clearly when $a=r_{h_2}$, $\dot a=0$ and for $a<r_{h_2}\,,\dot a>0$ and for $a>r_{h_2}\,,\dot a<0$. Thus, $a=r_{h_2}$ is a stable point which corresponds to the final static black hole. The above equation can be further solved analytically as
\be
g^2r_{h_2}(v-v_w)=\psi(a)-\psi(r_{h_1})\,,
\ee
where the function $\psi$ is defined by
\be \psi(a)=2\lambda\, \Big(\fft{a}{r_{h_2}} \Big)^{\fft{1}{\lambda}}\,{}_2F_1\Big(1\,,\ft{1}{\lambda+1}\,,\ft{\lambda+2}{\lambda+1}\,, \big(\fft{a}{r_{h_2}} \big)^{\fft{\lambda+1}{\lambda}}\Big) \,.\ee
When $v=v_w$, we have $\psi=\psi(r_{h_1})$, corresponding to the initial hairy black hole. Hence, $v=v_w$ can be interpreted as the initial time at which the shock wave is injected into the bulk.
\begin{figure}[ht]
\centering
\includegraphics[width=200pt]{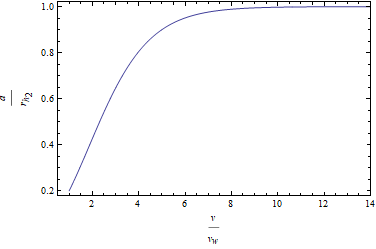}
\includegraphics[width=200pt]{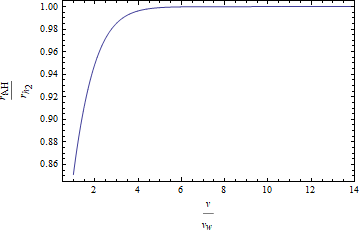}
\caption{{\it The ``scalar charge" $a$ (left panel) and the apparent horizon $r_{\mathrm{AH}}$ (right panel) are plotted as functions of the advanced time. We have set $\ell=1\,,r_{h_1}=1/5\,,\lambda=2$.}}
\label{scalarcharge}\end{figure}

The apparent horizon defined by $h\big(r_{\mathrm{AH}}(u)\big)=0$ can be solved as
\be r_{\mathrm{AH}}=\fft{1}{2}\Big(\sqrt{(\lambda+1)^2 a^2+8\lambda\ell^2\,\dot a}-(\lambda-1)a \Big) \,.\ee
At the stable point $a\rightarrow r_{h_2}\,,\dot a\rightarrow 0$, the apparent horizon approaches the event horizon of the final static black hole $r_{\mathrm{AH}}\rightarrow r_{h_2}$. In Fig.\ref{scalarcharge}, we show how the scalar charge and apparent horizon evolves in the dynamical process. It is easily seen that the apparent horizon evolves much faster than the scalar charge itself.

The Vaidya mass of the dynamical black hole is given by
\be M(v)=\fft{1}{16G\ell^2}\Big((\lambda+1)a^2+4\lambda\ell^2 \dot a \Big) \,.\ee
Hence, the injected energy of the shock wave reads
\bea
 \delta M(v)&=&M(v)-M_1 \nn\\
 &=&\fft{1}{16G\ell^2}\Big((\lambda+1)(a^2-r_{h_1}^2)+4\lambda\ell^2 \dot a \Big)\,.
 \eea
At the future infinity $v\rightarrow \infty$, the scalar charge exponentially approaches the stable limit. We find
\be
a(v)=r_{h_2}\Big(1-c_1 \,e^{-v/v_0}+\big(1-\ft{1}{2\lambda} \big)c_1^2\, e^{-2v/v_0}+\cdots \Big)\,,
\ee
where $c_1$ is a positive integration constant and $v_0$ is the relaxation time, given by
\be v_0=\fft{\lambda}{2\pi T_2} \,.\ee
It follows that
\bea
&&r_{\mathrm{AH}}=r_{h_2} \Big(1-\ft{\lambda-1}{2\lambda(\lambda+1)}\,c_1^2\,e^{-2v/v_0}+\cdots  \Big)  \,,\nn\\
&&\delta M=\Delta M\Big(1-\ft{(\lambda-1)M_2}{\lambda\Delta M}\,c_1^2\,e^{-2v/v_0}+\cdots  \Big)  \,,
\eea
where $\Delta M\equiv M_2-M_1$ is the total injected energy carried by the shock wave. It is clear that the apparent horizon and the injected energy approaches the stable limit much faster than the scalar charge. From these results, we conclude that the exact dynamical solution can be viewed as the collapse of a time-like shell with an effective finite width $\delta v\simeq v_0-v_w$.

\section{Action of the thin shell }

Considering a thin shell with a finite width $v_s-\epsilon\leq v\leq v_s+\epsilon$, we shall first identify the action on the shell in the thin shell limit. The shell separates the two static spacetimes, one is the black hole $v\geq v_s+\epsilon$ (called region 2) and the other is AdS vacuum $v\leq v_s-\epsilon$ (called region 1). The shell action reads
\begin{figure}
  \centering
  	\subfigure{\includegraphics[width=3in]{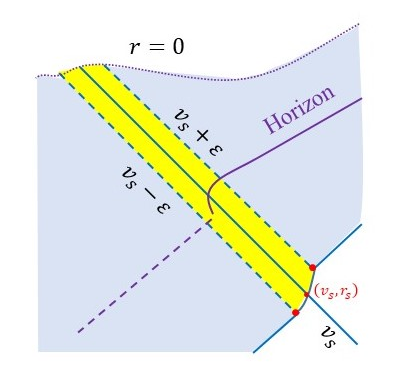}}
  	\caption{The collapse of a thin-shell.}
\label{fig2}\end{figure}
\bea\label{shellaction}
S_{shell}&=&\fft{1}{16\pi G}\int_{shell}\mathrm{d}^{n}x\,\sqrt{-g}\,\mathcal{L}+\fft{1}{8\pi G}\int_{r=0}\mathrm{d}^{n-1}x\,\sqrt{|\gamma|}\,K\nn\\
&&+\fft{1}{8\pi G}\int_{\mathcal{B}}d\lambda\mathrm{d}^{n-2}\theta\,\sqrt{|\gamma_N|}\,\kappa+\fft{1}{8\pi G}\int_{\Sigma}\mathrm{d}^{n-2}x\,\sqrt{\sigma}\,a\,,
\eea
where $\mathcal{B}$ denotes all the null boundaries. In the thin shell limit, the bulk volume vanishes so the bulk action does not have any contributions. Furthermore, the GH surface term at the future singularity vanishes as well in the limit $\epsilon\rightarrow 0$, as shown in \cite{CMM1}.

For the null boundaries, we impose normalizations $k\cdot \partial_t=\pm \alpha$ for the outward-directed normal vectors, where $\alpha$ is a positive constant and the sign ``$+(-)$" corresponds to the future (past) null boundaries. Thus, for the future null boundaries we can define
\bea
&&v=v_s+\epsilon\,,\quad k_+=\beta dv\,,\nn\\
&&v=v_s-\epsilon\,,\quad k_-=-\beta dv\,.
\eea
With this choice, it is clear that the above two null surfaces are affinely parameterized and hence the null surface term vanishes, namely $\kappa=0$. Therefore, we are left with only a portion of the past null boundary $\mathcal{B}_{past}$ and the two joints (denoted by the red points in the Fig.\ref{fig2}). The null norm of the past null boundary can be generally written as
\be\label{pastnullnorm} k^\mu \partial_\mu=H(r\,,v)\Big(\fft{2}{h(r\,,v)}\partial_v+\fft{1}{w(r\,,v)}\partial_r \Big) \,,\ee
or
\be k=H(r\,,v)\Big(-dv+\ft{2w(r\,,v)}{h(r\,,v)}dr \Big) \,,\ee
where $H(r\,,v)$ is a normalization function. We set
\bea
&&v>v_{max}=v_s+\epsilon\,,\quad H(r\,,v)=\alpha\,,\\\nn
&&v<v_{min}=v_s-\epsilon\,,\quad H(r\,,v)=\tilde\alpha\,,
\eea
where both $\alpha$ and $\tilde\alpha$ are positive constant but in general $\tilde\alpha\neq \alpha$ because the inner region does not extend to the asymptotic AdS boundary. In other words, we are freely to choose a proper $\tilde\alpha$ which will be very useful in the calculations of actions on the null boundaries. Using $k^\sigma\nabla_\sigma k^\mu=\kappa k^\mu$, we deduce
\bea
\kappa&=&k^\mu\partial_\mu \log{\Big(\fft{Hw}{h}\Big)}+k^r\partial_r\log{\Big( \fft hw \Big)} \nn\\
&=&\partial_\lambda \log{\Big(\fft{Hw}{h}\Big)}+\fft{H}{h}\partial_r\Big(\fft hw\Big)\,.
\eea
Thus,
\bea
S_{\mathcal{B}_{past}}&=&\fft{\omega_{n-2}}{8\pi G}\int d\lambda\, r^{n-2}\Big[ \partial_\lambda \log{\Big(\fft{Hw}{h}\Big)}+\fft{H}{h}\partial_r\Big(\fft hw\Big) \Big]\nn\\
&=& \fft{\omega_{n-2}}{8\pi G}\int d\lambda\, r^{n-2}\partial_\lambda \log{\Big(\fft{Hw}{h}\Big)}+\fft{\omega_{n-2}}{16\pi G}\int_{v_{min}}^{v_{max}} \mathrm{d}v\,r^{n-2}\partial_r\Big(\fft hw\Big)\,,
\eea
where in the second line we use the fact $k^v=\partial v/\partial \lambda=2H/h$. In the thin-shell limit, $r=r_s+\mathcal{O}(\epsilon/r_s)$. Thus to the leading order we have
\bea\label{pastaction}
\hat{S}_{\mathcal{B}_{past}}&=&\fft{\omega_{n-2}}{8\pi G} r_s^{n-2}\log{\Big(\fft{Hw}{h} \Big)}\Big|_{v_{min}\,,r_s}^{v_{max}\,,r_s} \nn\\
&=&\fft{\omega_{n-2}}{8\pi G} r_s^{n-2}\Big[\log{\Big(\fft{\alpha w_2(r_s)}{h_2(r_s)} \Big)}-\log{\Big(\fft{\tilde\alpha w_1(r_s)}{h_1(r_s)} \Big)} \Big]\,.
\eea
On the other hand, for the joint terms we have $a_\pm=\log{|k\cdot k_\pm|}$, giving rise to
\bea
&& a_+=\log{\Big(\fft{2\alpha\beta}{h_2(r_s)}\Big)}\,,\nn\\
&& a_-=\log{\Big(\fft{2\tilde\alpha\beta}{h_1(r_s)}\Big)}\,.
\eea
Following the prescriptions in \cite{CMM1}, we deduce
\bea\label{joint1}
\hat{S}_{joint}&=&\fft{1}{8\pi G}\int dS \big(a_--a_+ \big)\nn\\
&=&-\fft{\omega_{n-2}}{8\pi G} r_s^{n-2}\Big[\log{\Big(\fft{\alpha}{h_2(r_s)} \Big)}-\log{\Big(\fft{\tilde\alpha }{h_1(r_s)} \Big)} \Big]\,.
\eea
Combing the above results, we find in the thin shell limit
\be\label{finalshell} S_{shell}= \fft{\omega_{n-2}}{8\pi G} r_s^{n-2}\log{\Big( \fft{w_2(r_s)}{w_1(r_s)} \Big)}=\fft{\omega_{n-2}}{8\pi G} r_s^{n-2}\log{\Big( w_{BH}(r_s)\Big)} \,.\ee
In general, this term does not vanish except for the Vaidya-like metric which has $w_{BH}=1$.

Notice that this result does not depend on the normalization constants $\alpha\,,\widetilde\alpha$. In fact, the action on the past null boundary can be removed by choosing a proper normalization $\tilde\alpha$ to set $\kappa=0$. We easily find that such a $\tilde\alpha$ is given by
\be\label{tildealpha} \tilde\alpha=\fft{h_1(r_s)w_2(r_s)}{h_2(r_s)w_1(r_s)}\alpha=\fft{h_{vac}(r_s)w_{BH}(r_s)}{h_{BH}(r_s)}\alpha \,.\ee
which is also consistent to the result in \cite{CMM1}.
\section{ Boundary actions on WDW patch}\label{boundaryaction}
\textbf{GH surface contributions}\\
We now turn our attention to the boundary surface contributions in the action.
We first investigate the GH term at the past singularity. When $\tR<\tRc1$ or $\tL<\tLc1$,  the WDW patch intersects the past singularity and one finds
\bea\label{GHP}
S_{\GH}^{(p)}=-\lim_{r\rightarrow 0}\fft{\omega_{n-2}}{16\pi G}\,r^{n-2}w_{1}f_{1}\Big(\ft{h'_{1}}{h_{1}}+\ft{2(n-2)}{r}\Big)\big(-\tL+t_w+2\ro(r_s)-2\ro(0)\big) \,.
\eea
The time derivatives of this term are given by
\bea\label{GH1}
&&\frac{dS_{\GH}^{(p)}}{d\tR}=\lim_{r\rightarrow 0}\fft{\omega_{n-2}}{16\pi G}\,r^{n-2}w_{1}f_{1}\Big(\ft{h'_{1}}{h_{1}}+\ft{2(n-2)}{r}\Big)
\frac{w_2(r_s)f_2(r_s)}{w_1(r_s)f_1(r_s)}\,,\nn\\
&&\frac{dS_{\GH}^{(p)}}{d\tL}=\lim_{r\rightarrow 0}\fft{\omega_{n-2}}{16\pi G}\,r^{n-2}w_{1}f_{1}\Big(\ft{h'_{1}}{h_{1}}+\ft{2(n-2)}{r}\Big)\,.
\eea
Note that when $t_R>\tRc1\,,t_L>\tLc1$, this term does not exist since the WDW patch lifts off of the past singularity.

For the future singularity, when $\tL<\tLc2$, we have
\be\label{GHF}
S_{\GH}^{(f)}=-\lim_{r\rightarrow 0}\fft{\omega_{n-2}}{16\pi G}\,r^{n-2}w_{2}f_{2}\Big(\ft{h'_{2}}{h_{2}}+\ft{2(n-2)}{r}\Big)\big(\tR+t_w+2\rt(r_b)-2\rt(0)\big) \,,
\ee
which leads to
\bea\label{GH2}
&&\frac{dS_{\GH}^{(f)}}{d\tR}=-\lim_{r\rightarrow 0}\fft{\omega_{n-2}}{16\pi G}\,r^{n-2}w_{2}f_{2}\Big(\ft{h'_{2}}{h_{2}}+\ft{2(n-2)}{r}\Big)\,,\nn\\
&&\frac{dS_{\GH}^{(f)}}{d\tL}=-\lim_{r\rightarrow 0}\fft{\omega_{n-2}}{16\pi G}\,r^{n-2}w_{2}f_{2}\Big(\ft{h'_{2}}{h_{2}}+\ft{2(n-2)}{r}\Big)
\frac{w_1(r_b)f_1(r_b)}{w_2(r_b)f_2(r_b)}\,.
\eea
In contrast, when $\tL>\tLc2$, we have
\bea
S_{\GH}^{(f)}&=&-\lim_{r\rightarrow 0}\fft{\omega_{n-2}}{16\pi G}\,r^{n-2}w_{2}f_{2}\Big(\ft{h'_{2}}{h_{2}}+\ft{2(n-2)}{r}\Big)\big(\tR+t_w\big)\nn\\
&&-\lim_{r\rightarrow 0}\fft{\omega_{n-2}}{16\pi G}\,r^{n-2}w_{1}f_{1}\Big(\ft{h'_{1}}{h_{1}}+\ft{2(n-2)}{r}\Big)\big(-t_w+\tL-2\ro(0)\big)\,,
\eea
which results to
\bea\label{GH3}
&&\frac{dS_{\GH}^{(f)}}{d\tR}=-\lim_{r\rightarrow 0}\fft{\omega_{n-2}}{16\pi G}\,r^{n-2}w_{2}f_{2}\Big(\ft{h'_{2}}{h_{2}}+\ft{2(n-2)}{r}\Big)\,,\nn\\
&&\frac{dS_{\GH}^{f}}{d\tL}=-\lim_{r\rightarrow 0}\fft{\omega_{n-2}}{16\pi G}\,r^{n-2}w_{1}f_{1}\Big(\ft{h'_{1}}{h_{1}}+\ft{2(n-2)}{r}\Big)\,.
\eea
\\
\textbf{Joint contributions}\\
We move to the joint contributions to the action evaluated on the WDW patch. As shown in figure \ref{pert}, the only possible non-zero contributions to the time dependence of complexity come from the positions at $r=r_b, r_s$ and $r_m$.

We start our calculations from the joints, at which the null boundaries of the WDW patch get across the shock wave, $i.e.$,  $v=v_s$. The normal vectors along the collapsing shock wave is
\bea
&&v>-t_w\,,\quad k^s_+=-\beta dv\,,\nn\\
&&v<-t_w\,,\quad k^s_-=\beta dv\,.
\eea
For the joint at $r=r_s$, the normal vectors of the relevant null boundaries are given by
\be\label{kp}
k^{p}_{\mu} d x^{\mu}  =\left\lbrace \begin{matrix}
&\alpha \left( - d v + \frac{2}{w_2(r)f_{2}(r)} d r \right)&&
{\rm for}&& r>r_s\,,\\
&\tilde \alpha \left( - d v + \frac{2}{w_1(r)f_{1}(r)} d r \right)&&
{\rm for}&& r<r_s\,.
\end{matrix}\right.
\ee
It follows that its contribution to the action reads
\bea
S_{\joint}^{(\oor)}=\fft{\omega_{n-2}}{8\pi G} r_s^{n-2}\Big[\log{\Big(\fft{\alpha}{h_2(r_s)} \Big)}-\log{\Big(\fft{\tilde\alpha }{h_1(r_s)} \Big)} \Big]\,.
\eea
On the other hand, the (outward-directed) null norms to the future null boundary at $r=r_b$ is
\be\label{kf}
k^{f}_{\mu} d x^{\mu}  =\left\lbrace \begin{matrix}
&\alpha \left( - d v + \frac{2}{w_1(r)f_{1}(r)} d r \right)&&
{\rm for}&& r>r_b\,,\\
&\hat\alpha \left( - d v + \frac{2}{w_2(r)f_{2}(r)} d r \right)&&
{\rm for}&& r<r_b\,.
\end{matrix}\right.
\ee
Similarly, its action reads
\bea
S_{\joint}^{(\tr)}=\fft{\omega_{n-2}}{8\pi G} r_b^{n-2}\Big[\log{\Big(\fft{\alpha}{h_1(r_b)} \Big)}-\log{\Big(\fft{\hat\alpha }{h_2(r_b)} \Big)} \Big]\,.
\eea
As shown previously, we shall set $\kappa=0$ for the two past null boundaries by taking affine reparametrization condition
\be \fft{\tilde\alpha}{\alpha}=\fft{w_1(r_s)f_1(r_s)}{w_2(r_s)f_2(r_s)}\,,\qquad \fft{\hat\alpha}{\alpha}=\fft{w_2(r_b)f_2(r_b)}{w_1(r_b)f_1(r_b)} \,.\ee
However, as for the one-sided black hole case, these corner contributions are also exactly cancelled by the action of the shell, namely
\bea\label{jointrsrb}
S_{shell}+S_{\joint}^{(\oor)}+S_{\joint}^{(\tr)}=0\,.
\eea
Thus, we are left with the only possible nontrivial corner contribution coming from $r_m$. This joint contribution is evaluated with $k^p$ in Eq.(\ref{kp}) on the right boundary (with $r<r_s$) and
\be
k^L=\alpha dv\,,
\ee
for the normal vectors on the left null boundary. Here we have presumably assumed that the null normals are normalized asymptotically with the same constant at both the left and the right boundaries. The resulting joint contribution then reads,
\bea\label{jointiii}
S_{\joint}^{\ttr}=-\fft{\omega_{n-2}}{8\pi G} r_m^{n-2}\log{\Big(\fft{|h_1(r_m)|}{2\alpha\tilde\alpha} \Big)}=-\fft{\omega_{n-2}}{8\pi G} r_m^{n-2}\log{\Big(\fft{|h_1(r_m)|\,w_2(r_s)f_2(r_s)}{2\alpha^2\,w_1(r_s)f_1(r_s)} \Big)} \,,
\eea
where in the second ``=", we have picked out a factor $\ft{|h_1(r_m)|}{2\alpha^2}$, which equals to $|k^p\cdot k^L|$ before the shock wave turned on. In fact, since $\alpha$ is the asymptotic normalization of the null normals, it has nothing to do with the existence of shock waves and hence should be independent of $r_s(r_b)$. Thus, all the $r_s(r_b)$-dependence could be absorbed in $\tilde\alpha(\hat\alpha)$ owing to the affine reparametrization relation. It follows that the time derivatives of this joint contribution are given by
\bea\label{joint}
\frac{dS_{\joint}^{\ttr}}{d\tR}&=&\fft{\omega_{n-2}}{16\pi G} r_m^{n-2}\,w_1(r_m)f_1(r_m)\,\\
&&\Big[ \fft{h_1'(r_m)}{h_1(r_m)}
+\fft{n-2}{r_m}\log{\Big(\fft{|h_1(r_m)|\,w_2(r_s)f_2(r_s)}{2\alpha^2\,w_1(r_s)f_1(r_s)} \Big)}  \Big]\fft{w_2(r_s)f_2(r_s)}{w_1(r_s)f_1(r_s)}\nn\\
&&+\fft{\omega_{n-2}}{16\pi G} r_m^{n-2}\,w_2(r_s)f_2(r_s)\,\Big(\fft{f_2'(r_s)}{f_2(r_s)}+\fft{w_2'(r_s)}{w_2(r_s)}
-\fft{f_1'(r_s)}{f_1(r_s)}-\fft{w_1'(r_s)}{w_1(r_s)} \Big)  \,,\nn\\
\frac{dS_{\joint}^{\ttr}}{d\tL}&=&\fft{\omega_{n-2}}{16\pi G} r_m^{n-2}\,w_1(r_m)f_1(r_m)\,\Big[ \fft{h_1'(r_m)}{h_1(r_m)}+\fft{n-2}{r_m}\log{\Big(\fft{|h_1(r_m)|\,w_2(r_s)f_2(r_s)}{2\alpha^2\,w_1(r_s)f_1(r_s)} \Big)}  \Big]\,.\nn
\eea
It should be emphasized that these terms generally do not vanish even if when $\tL<\tLc1$ or $\tR<\tRc1$ where $r_m=0$.\\
\textbf{Counterterm contributions}

Next, we examine the contributions of the counterterm to the time evolution of holographic complexity. The discussions follow closely the section \ref{counteraction}.

For the past null boundary on the r.h.s of the WDW patch, we have
\bea\label{sct10}
S_{ct}^{\oor}&=&\fft{\omega_{n-2}}{8\pi G}\int_{r_{m}}^{r_s} \mathrm{d}\big(r^{n-2} \big)\log{\Big( \fft{(n-2)\ell_{ct} \tilde\alpha }{w_1\,r} \Big)}\nn\\
&&+\fft{\omega_{n-2}}{8\pi G}\int_{r_s}^{r_{max}} \mathrm{d}\big(r^{n-2} \big)\log{\Big( \fft{(n-2)\ell_{ct} \alpha }{w_{2}\, r} \Big)}\,,
\eea
 where we set the lower limit $r_{min}=r_m$, with the understanding that $r_m=0$ when the null boundaries end on the past singularity. To extract the time dependence, we deduce
\bea\label{sct1}
S_{ct}^{\oor}&=&\fft{\omega_{n-2}}{8\pi G}\Big\{ r_s^{n-2}\log{\Big(\fft{\tilde\alpha}{\alpha}\Big)}-\int^{r_s}_{r_m}\mathrm{d}\big( r^{n-2}\big)\log{w_1}\\
&&\qquad \quad-\int_{r_s}^{r_{max}}\mathrm{d}\big( r^{n-2}\big)\log{w_2}-r_m^{n-2}\Big[\log{\Big( \fft{(n-2)\ell_{ct}\tilde\alpha)}{ r_m}\Big)}+\fft{1}{n-2} \Big]+\cdots \Big\}\,,\nn
\eea
where the dots denotes the terms that are time independent.

For the future null boundary on the left side of the WDW patch, the counterterm action is given by
\bea\label{sct2}
S_{ct}^{\tr}&=&\fft{\omega_{n-2}}{8\pi G}\int_{r_b}^{r_{\max}} \mathrm{d}\big(r^{n-2} \big)\log{\Big( \fft{(n-2)\ell_{ct} \alpha }{w_1\,r} \Big)}\nn\\
&&+\fft{\omega_{n-2}}{8\pi G}\int_{0}^{r_b} \mathrm{d}\big(r^{n-2} \big)\log{\Big( \fft{(n-2)\ell_{ct} \hat\alpha }{w_{2}\, r} \Big)}\,,\nn\\
&=&\fft{\omega_{n-2}}{8\pi G}\Big[ r_b^{n-2}\log{\Big(\fft{\hat\alpha}{\alpha}\Big)}+\int^{r_b}\mathrm{d}\big( r^{n-2}\big)\log{\Big(\fft{w_1}{w_2}\Big)}+\cdots\Big]\,.
\eea
Here when $\tL>\tLc2$, we should drop this term since the null boundary no longer goes across the shock wave.

For the past null boundary extending to the left asymptotic AdS boundary, we have
\bea\label{sct3}
S_{ct}^{\ttr}&=&\fft{\omega_{n-2}}{8\pi G}\int_{r_{m}}^{r_{\max}} \mathrm{d}\big(r^{n-2} \big)\log{\Big( \fft{(n-2)\ell_{ct} \alpha }{w_1\,r} \Big)}\\
&=&\fft{\omega_{n-2}}{8\pi G}\Big\{\cdots-r_m^{n-2}\Big[\log{\Big( \fft{(n-2)\ell_{ct}\alpha)}{ r_m}\Big)}+\fft{1}{n-2} \Big]-\int^{r_{max}}_{r_m}\mathrm{d}\big( r^{n-2}\big)\log{w_1}\cdots \Big\}\,.\nn
\eea
where when $\tL<\tLc1$, the lower limit becomes $r_m=0$.

Finally, for the future null boundary on the right side of the WDW patch, we have
\bea\label{ctf}
S_{ct}^{\fr}&=&\fft{\omega_{n-2}}{8\pi G}\int_{0}^{r_{\max}} \mathrm{d}\big(r^{n-2} \big)\log{\Big( \fft{(n-2)\ell_{ct} \alpha }{w_2\,r} \Big)}\,,
\eea
which however does not depend on time.

To derive the time derivatives of these counterterm contributions, we begin with Eq.(\ref{sct2}). For the regime $\tL<\tLc2$, we have
\bea\label{sct2early}
&&\fft{dS_{ct}^{\tr}}{d\tR}=0\,,\nn\\
&&\fft{dS_{ct}^{\tr}}{d\tL}=\fft{\omega_{n-2}}{16\pi G}r_b^{n-2}\,w_1(r_b)f_1(r_b)\,\nn\\
&&\qquad\qquad\Big[\fft{f_2'(r_b)}{f_2(r_b)}+\fft{w_2'(r_b)}{w_2(r_b)}
-\fft{f_1'(r_b)}{f_1(r_b)}-\fft{w_1'(r_b)}{w_1(r_b)}+\fft{(n-2)}{r_b}\log{\Big( \fft{f_2(r_b)}{f_1(r_b)} \Big)} \Big]\,.
\eea
When $t_L>\tLc2$, the above terms vanish since $r_b$ meets the future singularity.

On the other hand, the time derivatives of $S_{ct}^{\oor}\,,S_{ct}^{\ttr}$ are deduced as follows. When $t_L<\tLc1$, $r_m=0$, we have
\bea\label{sct13early}
\fft{d}{dt_R}\big(S_{ct}^{\oor}+S_{ct}^{\ttr} \big)\Big|_{r_m=0}&=&\fft{\omega_{n-2}}{16\pi G}r_s^{n-2}
\,w_2(r_s)f_2(r_s)\,\\
&&\Big[ \fft{f_2'(r_s)}{f_2(r_s)}+\fft{w_2'(r_s)}{w_2(r_s)}-\fft{f_1'(r_s)}{f_1(r_s)}-\fft{w_1'(r_s)}{w_1(r_s)}
-\fft{(n-2)}{r_s}\log{\Big( \fft{f_1(r_s)}{f_2(r_s)}\Big)}  \Big]\,,\nn\\
\fft{d}{dt_L}\big(S_{ct}^{\oor}+S_{ct}^{\ttr} \big)\Big|_{r_m=0}&=&0\,,\nn
\eea
while when $t_L>\tLc1$, $r_m$ leaves the past singularity and we instead have
\bea\label{sct13late}
\fft{d}{dt_R}\big(S_{ct}^{\oor}+S_{ct}^{\ttr} \big)&=&\fft{d}{dt_R}\big(S_{ct}^{\oor}+S_{ct}^{\ttr} \big)\Big|_{r_m=0}\\
&&-\fft{\omega_{n-2}}{16\pi G}r_m^{n-2}\,w_2(r_s)f_2(r_s)
\Big\{ \fft{f_2'(r_s)}{f_2(r_s)}+\fft{w_2'(r_s)}{w_2(r_s)}-\fft{f_1'(r_s)}{f_1(r_s)}-\fft{w_1'(r_s)}{w_1(r_s)}\nn\\
&&-\fft{2(n-2)}{r_m} \fft{w_1(r_m)f_1(r_m)}{w_1(r_s)f_1(r_s)}
\log{\Big( \ft{(n-2)\ell_{ct}\alpha}{w_1(r_m)r_m}\ft{\sqrt{w_1(r_s)f_1(r_s)}}{\sqrt{w_2(r_s)f_2(r_s)}}  \Big)}\Big\}\,,\nn\\
\fft{d}{dt_L}\big(S_{ct}^{\oor}+S_{ct}^{\ttr} \big)&=&\fft{(n-2)\omega_{n-2}}{8\pi G}r_m^{n-3}\,w_1(r_m)f_1(r_m) \log{\Big( \ft{(n-2)\ell_{ct}\alpha}{w_1(r_m)r_m}\ft{\sqrt{w_1(r_s)f_1(r_s)}}{\sqrt{w_2(r_s)f_2(r_s)}}  \Big)}   \,.\nn
\eea

\section{Time dependence of complexity for eternal black holes}

Our derivation for complexity of shock wave geometries can be applied to the calculations of complexity for eternal black holes as well. In this case, the $r_s\,,r_b$ points (and the critical time $t_{L,c_2}$) are illusive so they do not have any time dependence. To derive correct results for eternal black holes, we shall take $h_1(r)=h_2(r)=h(r)\,,\cdots$. However, we should not include the results depending on $t_L<\tLc2$ since this critical time does not truly exist. Thus, we should drop the GH surface term at the future singularity Eq.(\ref{GH2}) and the counterterm Eq.(\ref{sct2early}). By taking $t_L=t_R=t/2$, we deduce
(dropping Eq.(\ref{tc21}))
\bea
&&\fft{dS_{bulk}}{dt}=\fft{\omega_{n-2}}{16\pi G}\int_0^{r_m}\mathrm{d}r\,\sqrt{-\bar g}\,\mathcal{L} =\fft{\omega_{n-2}}{16\pi G} \Big(\epsilon^{n-2}\fft{h'(\epsilon)}{w(\epsilon)}-r_m^{n-2}\fft{h'(r_m)}{w(r_m)}  \Big)     \,,\nn\\
&&\fft{dS_{GH}}{dt}=-\fft{\omega_{n-2}}{16\pi G}\epsilon^{n-2}w(\epsilon)f(\epsilon)\Big(\fft{h'(\epsilon)}{h(\epsilon)}+\fft{2(n-2)}{\epsilon}  \Big)        \,,\nn\\
&&\fft{dS_{joint}}{dt}=\fft{\omega_{n-2}}{16\pi G}r_m^{n-2}w(r_m)f(r_m)\Big(\fft{h'(r_m)}{h(r_m)}+\fft{n-2}{r_m}\log{\Big( \fft{|h(r_m)|}{2\alpha^2} \Big) }  \Big)       \,,
\eea
where it was understood that $\epsilon\rightarrow 0$. Combining the above results, we obtain
\be
\fft{dS}{dt}=-\fft{(n-2)\omega_{n-2}}{8\pi G}\epsilon^{n-3}w(\epsilon)f(\epsilon)+\fft{(n-2)\omega_{n-2}}{16\pi G}r_m^{n-3}w(r_m)f(r_m)\log{\Big( \fft{|h(r_m)|}{2\alpha^2} \Big) }\,,
\ee
where the first term on the r.h.s is the correct late time limit of the action growth (for eternal black holes in Einstein-Scalar gravity). Of course, this is consistent with the above formula since in that limit, $r_m\rightarrow r_h\,,f(r_m)\,,h(r_m)\rightarrow 0$. The above result can be applied to the CA-2 and CV-2 duality as well. One has
\bea
&&\fft{dS_{\Lambda}}{dt}=-\fft{\omega_{n-2}}{16\pi G}\int_0^{r_m}\mathrm{d}r\,\sqrt{-\bar g}\,V_\Lambda\,,\nn\\
&&\fft{dS_{V}}{dt}=-\fft{\Lambda\omega_{n-2}}{8\pi G}\int_0^{r_m}\mathrm{d}r\,\sqrt{-\bar g}\,.
\eea
Without shock waves, the time dependence of complexity is fully determined by the evolution of the position $r_m$. One easily finds
\be t_R+t_L=-2r^*(r_m) \,,\ee
which depends only on the combination $t_R+t_L$. This is a reminiscent of the symmetry of the problem. The action is invariant under the time translations $t_R\rightarrow t_R+\delta t\,,t_L\rightarrow t_L-\delta t$. It follows that the critical time where the position $r_m$ lifts off of the past singularity is given by $t_c=-2r^*(0)$. For $t<t_c$, the rate of change of complexity is a constant while for $t>t_c$ it grows nonlinearly with time and approaches the late time limit at $t\rightarrow \infty$.

\end{document}